\documentclass[aps,prb,twocolumn,superscriptaddress,longbibliography]{revtex4-2}

\usepackage{amssymb,amsmath,amsthm,bbm,bbold}
\usepackage{amsfonts}
\usepackage{graphicx}
\usepackage{mathtools}
\usepackage[breaklinks,colorlinks,allcolors=blue]{hyperref}
\usepackage[normalem]{ulem}
\usepackage[T1]{fontenc}
\usepackage{xcolor}
\usepackage{tcolorbox}

\DeclareMathOperator{\dist}{dist}
\usepackage{longtable}
\usepackage{tikz}

\usepackage{makecell}

\def\RR{\mathbbm{R}}
\def\CC{\mathbbm{C}}

\newcommand{\Tr}{\mathrm{Tr}}

\newcommand{\bra}[1]{\mbox{$\langle #1 |$}}
\newcommand{\ket}[1]{\mbox{$| #1 \rangle$}}
\newcommand{\bd}[1]{\boldsymbol{#1}}
\newcommand{\bl}{\boldsymbol{\lambda}}

\newcommand*\xbar[1]{\hbox{\vbox{
       \hrule height 0.6pt % The actual bar
       \kern0.3ex%         % Distance between bar and symbol
       \hbox{%
         \kern-0.2em%      % Shortening on the left side
         \ensuremath{#1}%
         \kern 0.0em%      % Shortening on the right side
         }}}}

\newcommand*\xxbar[1]{\hbox{\vbox{
       \hrule height 0.6pt % The actual bar
       \kern0.3ex%         % Distance between bar and symbol
       \hbox{%
         \kern-0.0em%      % Shortening on the left side
         \ensuremath{#1}%
         \kern 0.0em%      % Shortening on the right side
         }}}}

\begin{document}
\title{A toolbox of spin-adapted generalized Pauli constraints
}

\author{Julia Liebert}
\affiliation{%
	Department of Physics, Arnold Sommerfeld Center for Theoretical Physics,
	Ludwig-Maximilians-Universit\"at M\"unchen,
	Theresienstr.~37, 80333 Munich, Germany
}
\affiliation{%
	Munich Center for Quantum Science and Technology (MCQST),
	Schellingstr.~4, 80799 Munich, Germany
}

\author{Yannick Lemke}
\affiliation{%
	Chair of Theoretical Chemistry, Department of Chemistry,
	Ludwig-Maximilians-Universit\"at M\"unchen,
	Butenandtstr.~5--13, 81377 Munich, Germany
}

\author{Murat Altunbulak}
\affiliation{%
	Department of Mathematics, Faculty of Science,
	Dokuz Eylul University, 35390 Buca-Izmir, Turkey}

\author{Tomasz Maciazek}
\affiliation{School of Mathematics, University of Bristol, Fry Building, Woodland Road, Bristol, BS8 1UG, United Kingdom}

\author{Christian Ochsenfeld}
\affiliation{%
	Chair of Theoretical Chemistry, Department of Chemistry,
	Ludwig-Maximilians-Universit\"at M\"unchen,
	Butenandtstr.~5--13, 81377 Munich, Germany
}
\affiliation{%
	Max-Planck-Institute for Solid State Research,
	Heisenbergstr.~1, 70569 Stuttgart, Germany
}

\author{Christian Schilling}
\email{c.schilling@physik.uni-muenchen.de}
\affiliation{%
	Department of Physics, Arnold Sommerfeld Center for Theoretical Physics,
	Ludwig-Maximilians-Universit\"at M\"unchen,
	Theresienstr.~37, 80333 Munich, Germany
}
\affiliation{%
	Munich Center for Quantum Science and Technology (MCQST),
	Schellingstr.~4, 80799 Munich, Germany
}

\date{\today}

\begin{abstract}
We establish a toolbox for studying and applying spin-adapted generalized Pauli constraints (GPCs) in few-electron quantum systems. By exploiting the spin symmetry of realistic $N$-electron wave functions, the underlying one-body pure $N$-representability problem simplifies, allowing us to calculate the GPCs for larger system sizes than previously accessible. We then uncover and rigorously prove a superselection rule that highlights the significance of GPCs: whenever a spin-adapted GPC is (approximately) saturated --- referred to as (quasi)pinning --- the corresponding $N$-electron wave function assumes a simplified structure. Specifically, in a configuration interaction expansion based on natural orbitals only very specific spin configuration state functions may contribute. To assess the nontriviality of (quasi)pinning, we introduce a geometric measure that contrasts it with the (quasi)pinning induced by simple (spin-adapted) Pauli constraints. Applications to few-electron systems suggest that previously observed quasipinning largely stems from spin symmetries.
\end{abstract}

\maketitle

\section{Introduction \label{sec:intro}}

Pauli's \emph{exclusion principle} \cite{Pauli1925} is fundamental to our understanding of fermionic quantum systems, governing phenomena across all length scales, from subatomic particles to neutron stars. It ensures the stability of matter \cite{Lieb1976, Lieb-book} and underlies the \emph{Aufbau principle}, which dictates atomic structure and the periodic table. Beyond the Pauli principle, fermionic exchange symmetry imposes additional constraints on occupation numbers \cite{BD72,R07}. However, a complete classification of these \emph{generalized Pauli constraints} (GPCs) remained elusive for decades until a systematic framework was developed \cite{Klyachko2004, K06, AK08}. These constraints take the form of linear inequalities on the decreasingly ordered \emph{natural occupation numbers} $\lambda_j$, the eigenvalues of the one-particle reduced density matrix. For each particle number $N$ and one-particle Hilbert space dimension $d$, the GPCs define a convex polytope in $\mathbbm{R}^d$ of admissible occupation number vectors $\bd\lambda$, which is contained within the \emph{Pauli simplex} $1\geq \lambda_1\geq \dots \geq \lambda_d\geq 0$.

To assess their physical relevance, GPCs have been studied in various few-fermion systems through analytical \cite{SGC13, Schilling2014, BenavidesRiveros2015, S15-H, TEVS16, TVS16, CBV17, BM19, MSGLS20, SBLMS20, Reuvers2021, Hackl2023} and numerical approaches \cite{BGS13, Chakraborty2014, CM15, Chakraborty2015, CM16-ES, DePrince16, Chakraborty2017, LS18, TLH18, SAKL18, Chakraborty2018, Boyn2019, Smart2019, Avdic2023, ASGWHEM23}. Initially, it was conjectured that ground-state occupation numbers might exactly saturate some GPCs, a phenomenon termed \emph{pinning} \cite{K09}. However, this was later refuted when sufficiently accurate wave function approximations were employed (see, e.g., \cite{SGC13, SAKL18, LS18}). Instead, \emph{quasipinning} was observed, meaning occupation number vectors lie remarkably close to but not exactly on the polytope boundary \cite{SGC13, BGS13, S15-H, CM15, TEVS16, TVS16, CBV17, LS18, SAKL18}.

The presence of (quasi)pinning has important consequences: it constrains the system's response to external perturbations \cite{K09,Klyachko2013,S15-H} and implies a significantly simplified structure of the $N$-fermion wave function \cite{Klyachko2013,CBV17, MSGLS20, SBLMS20}. This insight led to a hierarchy of multi-configurational self-consistent field (MCSCF) variational \emph{ans\"atze} \cite{BenavidesRiveros2016,SBLMS20}, systematically incorporating electron correlation by relating polytope faces of increasing dimensionality. Furthermore, GPCs play a crucial role in one-particle reduced density matrix functional theory (RDMFT), where they define the domain of the pure-state functional and strongly influence its general form \cite{Schilling18, LCLS21, LCS23}. Notably, RDMFT has shown that fermionic exchange symmetry manifests in the one-particle picture as a repulsive \emph{kinematical exchange force}, which diverges at the polytope boundary. This, in turn, provides a comprehensive explanation for the absence of exact pinning in realistic systems \cite{CS19ExForce,LCS23}.

Despite their potential significance, the application of GPCs in physics and chemistry remains limited due to their complexity and the lack of efficient algorithms for their calculation. Currently, GPCs are explicitly known only for systems with up to ten spin orbitals \cite{AK08, Altunbulak-phd} and, for $N=3$, up to twelve spin orbitals \cite{SAKL18}.

This work addresses these limitations by extending previous computational efforts to incorporate spin symmetries in realistic electronic systems. Specifically, we determine spin-adapted GPCs for larger active spaces than previously accessible, reaching up to eight spatial orbitals (16 spin orbitals). Moreover, we establish a practical framework for applying spin-adapted GPCs, providing a systematic approach to exploring quasipinning in realistic fermionic systems while circumventing common pitfalls.

Our paper is organized as follows. Section~\ref{sec:notation} introduces key concepts, while Sec.~\ref{sec:spin-gpcs} examines the implications of spin symmetries within the one-body $N$-representability problem, leading to spin-adapted GPCs. We then demonstrate constructively that (approximate) saturation of these constraints implies a significantly simplified structure of the corresponding $N$-electron wave function, with applications in wave function theory and RDMFT. Section~\ref{sec:BD} provides an illustrative example. To reliably distinguish genuine quasipinning from trivial cases arising due to weak or reduced electron correlation, we introduce the $Q$-parameter in Sec.~\ref{sec:Q-param}, a geometric quantifier of non-triviality. Finally, applying our toolbox to few-electron atoms in Sec.~\ref{sec:example}, we show that previously observed (quasi)pinning in atomic systems \cite{SAKL18} largely originates from spin symmetries.

\section{Key concepts and notation\label{sec:notation}}

\subsection{Pure state $N$-representability problem \label{sec:Nrep}}

The one-particle Hilbert space of a spin-$1/2$ fermion is given by
\begin{equation}
\mathcal{H}_1 \equiv \mathcal{H}_1^{(l)}\otimes \mathcal{H}_1^{(s)}\,,
\end{equation}
where $\mathcal{H}_1^{(l)}\cong \CC^d$ represents the orbital component and $\mathcal{H}_1^{(s)}\cong\CC^2$ the spin component, with a finite orbital dimension $d = \mathrm{dim}(\mathcal{H}_1^{(l)})$. The corresponding Hilbert space for $N$ fermions follows as the antisymmetrized $N$-fold product,
\begin{equation}
\mathcal{H}_N = \wedge^N\mathcal{H}_1\,.
\end{equation}
The quantum states of $N$ fermions are then described by density matrices on $\mathcal{H}_N$, with pure states given by projectors $\Gamma \equiv\ket{\Psi}\!\bra{\Psi}$, where $\ket{\Psi}\in \mathcal{H}_N$ is normalized. The set of all pure-state density matrices is denoted by
\begin{equation}\label{eq:PN}
\mathcal{P}^N :=  \left\{\Gamma\,|\,\Gamma^2=\Gamma, \Gamma\geq 0, \Tr_N[\Gamma]=1\right\}\,.
\end{equation}

To each $N$-fermion state $\Gamma$, we associate its one-particle reduced density matrix (1RDM), obtained by tracing out $N-1$ particles,
\begin{equation}\label{eq:1RDM}
\gamma:= N\Tr_{N-1}[\Gamma] \equiv \sum_{j=1}^{2d}\lambda_j \ket{j}\bra{j}\,,
\end{equation}
where $\lambda_j$ and $\ket{j}$ are the \emph{natural occupation numbers} and \emph{natural spin-orbitals}, respectively.

The one-body pure $N$-representability problem aims to characterize the set
\begin{equation}\label{eq:P1N}
\mathcal{P}^1_N:= \left\{\gamma\,|\,\exists\Gamma\in\mathcal{P}^N\!\!: N\Tr_{N-1}[\Gamma]=\gamma\right\},
\end{equation}
which contains all admissible 1RDMs. A fundamental property of $\mathcal{P}^1_N$ is its invariance under unitary transformations: for any $\gamma \in \mathcal{P}^1_N$ and any unitary $u$ on $\mathcal{H}_1$, the transformed matrix $\tilde{\gamma}\equiv u \gamma u^\dagger$ also belongs to $\mathcal{P}^1_N$ (see, e.g., Ref.~\cite{Schilling2014}). Consequently, the constraints defining $\mathcal{P}^1_N$ involve only the natural occupation numbers.

Building upon this observation, Klyachko and Altunbulak \cite{C63, K06, AK08} established that the admissible decreasingly-ordered natural occupation number vectors $\bd\lambda \equiv \bd\lambda^\downarrow$ form a convex polytope. They further developed a systematic framework to determine the corresponding facet-defining inequalities. Each of these finitely-many \textit{generalized Pauli constraints} (GPCs) takes the form
\begin{equation}\label{eq:gpc}
D(\bd \lambda) =\kappa_0+\sum_{j=1}^{2d}\kappa_j\lambda_j\geq 0\,,
\end{equation}
with distinctive integer coefficients $\kappa_j$.

A natural occupation number vector $\bd\lambda$ is said to be \emph{pinned} by a GPC $D \geq 0$ if $D(\bd\lambda)=0$ \cite{K09} and \emph{quasipinned} if the equality holds approximately \cite{SGC13,Schilling2014}. Pinning has significant physical consequences: it restricts the system's linear response to external perturbations \cite{K09,Klyachko2013,S15-H} and enforces a simplified structure of the $N$-fermion wave function \cite{Klyachko2013,CBV17, MSGLS20, SBLMS20}.

To recall this, for a given state $\ket{\Psi}$ each GPC $D \geq 0$ can be associated with a Hermitian one-particle operator on $\mathcal{H}_N$,
\begin{equation}\label{eq:D-hat}
\hat{D}^{(\Psi)}:= \kappa_0 \mathbbm{1} +\sum_{j=1}^{2d}\kappa_j \hat{n}_j\,.
\end{equation}
Here, $\hat{n}_j = f_{j}^\dagger f_j$ is the occupation number operator, with $f_j^\dagger$ and $f_j$ denoting the fermionic creation and annihilation operators for spin-orbital $\ket{j}$. The dependence of $\hat{D}^{(\Psi)}$ on $\ket{\Psi}$ through its spin-orbitals is made explicit, but for simplicity, superscripts are often omitted. If $D(\bd\lambda)=0$, then \cite{Klyachko2013,CBV17, MSGLS20, SBLMS20}
\begin{equation}\label{eq:DtoPsi}
\hat{D}^{(\Psi)}\ket{\Psi}=0\,.
\end{equation}
Since the eigenstates of $\hat{D}^{(\Psi)}$ are occupation number states $\ket{\bd j} = f_{j_1}^\dagger \ldots f_{j_N}^\dagger\ket{0}$, this implies a \emph{superselection rule} restricting the contributing configurations in the wave function expansion,
\begin{equation}\label{eq:Psi-selec}
\ket{\Psi} = \sum_{\bd j\in\mathcal{I}_N^{(D)}} c_{\bd j}\ket{\bd j}\,.
\end{equation}
The index set $\mathcal{I}_N^{(D)}$ contains precisely those configurations $\bd j$ for which $\hat{D}^{(\Psi)}\ket{\bd j}=0$ \cite{K09, Schilling15, CBV17}. This selection rule is stable in the sense that quasipinning ($D(\bd\lambda) \approx 0$) implies an approximate restriction of the wave function structure \cite{Schilling15,CBV17}. This result suggests to employ \eqref{eq:Psi-selec}, and also more restricted expansions corresponding to (quasi)pinning by multiple GPCs, as multi-configurational self-consistent field (MCSCF) ans\"atze, as worked out in \cite{TVS17, SBLMS20}.

The presence of (quasi)pinning can thus be understood as a consequence of a formal $U(1)$ symmetry generated by $\hat{D}^{(\Psi)}$. This perspective explains why symmetries increase the likelihood of (quasi)pinning and, in some cases, enforce it entirely. A key objective of this work is to investigate to what extent the (quasi)pinning previously observed in few-electron atoms arises from (spin-)symmetries. To achieve this, we integrate symmetries directly into the GPC framework, ensuring a fully consistent and symmetry-adapted formalism.

\subsection{Spin symmetries\label{sec:spin-symm}}

Spin $\text{SU}(2)$ symmetry is a fundamental property of physical Hamiltonians. In this section, we first review the general formalism of spin symmetries before examining how incorporating electron spin into the GPC framework affects (quasi)pinning.

For spin-$1/2$ fermions, the generators of the Lie algebra $\mathfrak{su}(2)$ of the non-abelian Lie group $\text{SU}(2)$ are given by
\begin{equation}\label{eq:Si}
S^\nu_j =\frac{1}{2}\sum_{\sigma, \sigma^\prime}f_{j\sigma}^\dagger(\bd\tau_\nu)_{\sigma\sigma^\prime}f_{j\sigma^\prime}, \quad \nu = x,y,z,
\end{equation}
and for the total system of $d$ orbitals by $S^\nu = \sum_{j=1}^{d}S^\nu_j$. Here, $f_{j\sigma}^\dagger$ and $f_{j\sigma}$ create and annihilate fermions in the spin-orbital $\ket{j\sigma}$, respectively, while $\bd\tau = (\tau_x, \tau_y, \tau_z)$ denotes the Pauli matrices
\begin{equation}
\tau_x = \begin{pmatrix}
0&1\\ 1&0
\end{pmatrix}, \quad
\tau_y = \begin{pmatrix}
0&-\mathrm{i}\\ \mathrm{i}&0
\end{pmatrix}, \quad
\tau_z = \begin{pmatrix}
1&0\\ 0&-1
\end{pmatrix}.
\end{equation}
The total spin operator $\bd S^2$ is a Casimir operator of $\text{SU}(2)$ and as such commutes with all three generators $S^x, S^y, S^z$. In second quantization, it takes the form
\begin{equation}
\bd S^2 =  \sum_{i,j=1}^{d}S_i^zS^z_j+\frac{1}{2}\left(S^+_iS^-_j + S^-_iS^+_j\right),
\end{equation}
where $S^\pm$ are the spin raising and lowering operators,
\begin{equation}\label{eq:Spm}
S^{\pm}_j:= S^x_j\pm \mathrm{i} S^y_j.
\end{equation}
The total spin quantum number $S$ and total magnetization $M$ follow from the eigenvalue equations $\bd S^2\ket{\Psi} = S(S+1)\ket{\Psi}$ and $S_z\ket{\Psi} = M\ket{\Psi}$ for $\ket{\Psi}\in\mathcal{H}_N^{(S,M)}$.

We now examine the expansion of $N$-electron states $\ket{\Psi}$ in spin-configuration states, adapting the standard expansion in Slater determinants $\ket{\boldsymbol{j}}$ to incorporate spin symmetry. This will be crucial in Sec.~\ref{sec:selection} for formulating an analogous superselection rule when spin-adapted GPCs are (quasi)pinned.

By the Peter--Weyl theorem \cite{Hall15}, the $N$-fermion Hilbert space decomposes into irreducible unitary representations labeled by $S$,
\begin{equation}
\mathcal{H}_N = \bigoplus_{S =S_{\min}}^{S_{\max}}\mathcal{H}_N^{(S)},
\end{equation}
where $S_{\max}=N/2$ and $S_{\min}=0$ for $N$ even, while $S_{\min}=1/2$ for $N$ odd. Each irreducible representation $\mathcal{H}_N^{(S)}=\oplus_{M=-S}^S\mathcal{H}_N^{(S,M)}$ contains a \emph{unique} highest weight state $\ket{\Lambda_S}$ satisfying \cite{Verma68, LMNP86, RS99, Hall15, PP18}
\begin{equation}\label{eq:condition-hw-S}
\forall j<i\!:\,E_{ij}\ket{\Lambda_{S}} = 0\quad \wedge\quad S^{+}\ket{\Lambda_{S}} = 0\,,
\end{equation}
where
\begin{equation}\label{eq:Eij}
E_{ij} := f_{j\uparrow}^\dagger f_{i\uparrow} +f_{j\downarrow}^\dagger f_{i\downarrow}\,.
\end{equation}
The highest weight state for $\mathcal{H}_N^{(S)}$ is
\begin{equation}\label{eq:Lambda-hwv}
\ket{\Lambda_{S}} = \ket{\underbrace{1\!\uparrow, 1\!\downarrow, \ldots, K\!\uparrow, K\!\downarrow}_{N-2S}, \underbrace{(K+1)\!\uparrow, \ldots, J\!\uparrow}_{2S}},
\end{equation}
where
\begin{equation}\label{eq:kj}
K=\frac{N-2S}{2}, \quad J=\frac{N+2S}{2}.
\end{equation}
A monomial basis (Verma basis) \cite{Verma68, LMNP86, RS99, Hall15, PP18} for $\mathcal{H}_N^{(S)}$ is constructed by successively applying the negative root operators $E_{ij}^\dagger$ ($i=j+1$) and $S^-$ to $\ket{\Lambda_S}$. This generates a set of spin-adapted configuration states $\ket{I}$, indexed by $\mathcal{I}_{N,S}$, analogous to the spin-independent configurations $\bd{j}=(j_1,\ldots,j_N)$ in Sec.~\ref{sec:Nrep}. Any state $\ket{\Psi} \in \mathcal{H}_N^{(S)}$ can then be expanded as
\begin{equation}\label{eq:PsiI}
\ket{\Psi} = \sum_{I \in \mathcal{I}_{N,S}} c_I \ket{I}\,.
\end{equation}

Similarly, for each symmetry sector $\mathcal{H}_N^{(S,M)}$, the highest weight state is given by \cite{Verma68, LMNP86, RS99, Hall15, PP18}
\begin{equation}\label{eq:Lambda-SM}
\ket{\Lambda_{S,M}} = \frac{(S^-)^{S-M} \ket{\Lambda_S}}{|| (S^-)^{S-M} \ket{\Lambda_S}||_2},
\end{equation}
while the basis states $\{\ket{I}\}_{I \in \mathcal{I}_{N,S,M}}$ spanning $\mathcal{H}_N^{(S,M)}$ are constructed via monomials of $E_{j+1,j}^\dagger$ ($j=1,\ldots,d-1$). A basis for $\mathcal{H}_N^{(S,M)}$ can also be obtained by selecting from a basis of $\mathcal{H}_N^{(S)}$ only those states $\ket{I}$ with magnetization $M$. Thus, any state $\ket{\Psi} \in \mathcal{H}_N^{(S,M)}$ can be expanded as in Eq.~\eqref{eq:PsiI}, with configurations restricted to $\mathcal{I}_{N,S,M}$,
\begin{equation}\label{eq:PsiISM}
\ket{\Psi} = \sum_{I \in \mathcal{I}_{N,S,M}} c_I \ket{I}.
\end{equation}

In analogy to Eqs.~\eqref{eq:PN} and \eqref{eq:P1N}, the set $\mathcal{P}^N_{S, M}$ consists of pure states $\Gamma = \ket{\Psi}\!\bra{\Psi}$ with $\ket{\Psi}\in\mathcal{H}_N^{(S,M)}$. The set of \textit{one-body pure $(N,S,M)$-representable} 1RDMs is
\begin{equation}\label{eq:P1NSM}
\mathcal{P}^1_{N,S,M}:= N\Tr_{N-1}\left[\mathcal{P}^N_{S, M}\right]\,.
\end{equation}
Both $\mathcal{P}^N_{S,M}$ and $\mathcal{P}^1_{N,S,M}$ are generally non-convex. Since $M$ is conserved, any 1RDM $\gamma\in \mathcal{P}^1_{N,S,M}$ is block-diagonal in the spin basis,
\begin{equation}
\gamma = \gamma^{\uparrow\uparrow}\oplus \gamma^{\downarrow\downarrow}, \quad \gamma_{ij}^{\sigma\sigma} =\mathrm{Tr}_1[f_{j\sigma}^\dagger f_{i\sigma}]\,.
\end{equation}

All the concepts introduced in this section will be essential for analyzing the structural implications of (quasi)pinning in the following parts of this work.

\section{Spin symmetry-adapted generalized Pauli constraints \label{sec:spin-gpcs}}

This section introduces spin-adapted generalized Pauli constraints, proving and illustrating how the structure of the corresponding $N$-fermion wave function simplifies in the presence of (quasi)pinning.

\subsection{Spin-adapted orbital one-body $N$-representability problem\label{sec:orbital}}

In second quantization, the generators of the unitary group $\text{U}(d)\subset \text{U}(2d)$ describing orbital rotations are given by the operators $E_{ij}$ defined in Eq.~\eqref{eq:Eij}. Similarly, the generators $S_{\sigma, \sigma^\prime}$ of $\text{U}(2)$ arise from the $\text{U}(2d)$ generators $f_{j\sigma^\prime}^\dagger f_{i\sigma}$ by summing over the orbital degrees of freedom:
\begin{equation}
S_{\sigma, \sigma^\prime}:= \sum_{i=1}^d f_{i\sigma^\prime}^\dagger f_{i\sigma}\,.
\end{equation}
These relate to the spin raising/lowering operators in Eq.~\eqref{eq:Spm} via $S^+ = S_{\downarrow\uparrow}$ and $S^- = S_{\uparrow\downarrow}$.

The orbital 1RDM in a chosen basis $\mathcal{B}_l$ of $\mathcal{H}_1^{(l)}$ follows directly from the $\text{U}(d)$ generators (recall \eqref{eq:Eij}):
\begin{equation}\label{eq:gl-ij}
(\gamma_l)_{ij} = \Tr[E_{ij}\Gamma]\,,
\end{equation}
where $\Gamma$ is an $N$-fermion quantum state, either pure or mixed. Since the $E_{ij}$ operators commute with $S^\pm$, the orbital 1RDM $\gamma_l$ remains the same for all states of a given spin multiplet.

In analogy to $\mathcal{P}^1_{N,S,M}$ in Eq.~\eqref{eq:P1NSM}, we define the set of all orbital 1RDMs that can arise from a pure $N$-fermion state $\Gamma\in \mathcal{P}^N_{S,M}$ as
\begin{equation}
\mathcal{L}^1_{N,S,M}:= \Big\{\gamma_l\,\Big\vert\, \exists \Gamma\in\mathcal{P}^N_{S,M} \text{ such that } \Gamma\mapsto\gamma_l \Big\}\,.
\end{equation}
Since the partial trace $\Gamma\mapsto\gamma_l$ is linear, the convex hull of this set defines the admissible $\gamma_l$ for ensemble states,
\begin{equation}
\xxbar{\mathcal{L}}^1_{N,S,M} = \mathrm{conv}\left(\mathcal{L}^1_{N,S,M}\right)\,.
\end{equation}
In striking contrast to the set \eqref{eq:P1NSM} of admissible full 1RDMs, both $\mathcal{L}^1_{N,S,M}$ and $\xxbar{\mathcal{L}}^1_{N,S,M}$ are invariant under orbital rotations $u_l$.
Consequently, $\mathcal{L}^1_{N,S,M}$ is fully characterized by the spectral set
\begin{equation}
\Sigma^{(p)\downarrow}_{S,M}:= \left\{\bd\lambda^{(l)}\,\Big\vert\,\exists \gamma_l\in \mathcal{L}^1_{N,S,M} \text{ such that } \bd\lambda^{(l)} = \mathrm{spec}^\downarrow(\gamma_l)\right\}\,
\end{equation}
similar to the spectral polytope described by the spin-independent GPCs for the full 1RDM $\gamma$. The polytope $\Sigma^{(p)\downarrow}_{S,M}$ is described by the \textit{spin-adapted generalized Pauli constraints} \cite{AK08, Altunbulak-phd}, where $\lambda_i^{(l)}$ denote the eigenvalues of $\gamma_l$, known as \emph{natural orbital occupation numbers}, and the corresponding eigenstates $\ket{i}$ are the \emph{natural orbitals}.
\begin{figure*}[tb]
\includegraphics[width=\linewidth]{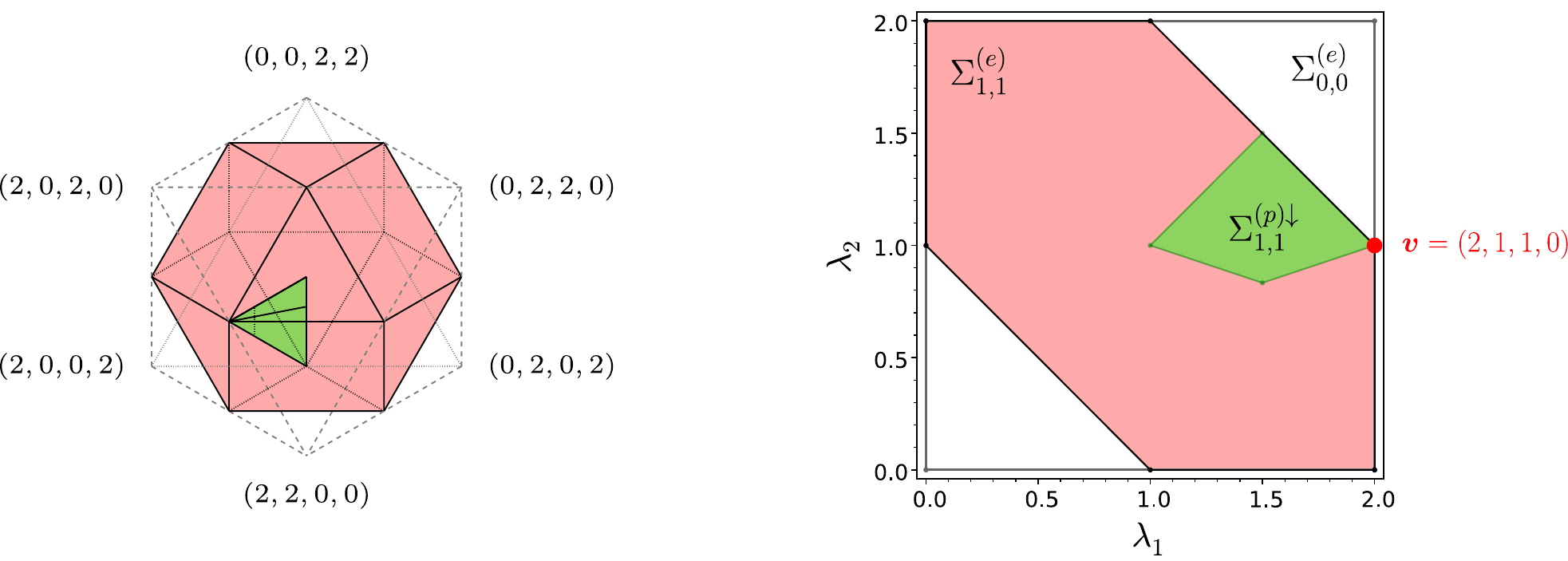}
\caption{Illustration of the spin-adapted Pauli polytope $\Sigma^{(e)}_{S,M}$ \eqref{eq:PC_SM} and the polytope $\Sigma^{(p)\downarrow}_{S,M}$ characterized by the spin-adapted GPCs for $N=d=4, S=1$ \eqref{eq:GPC_31}. Left: Three-dimensional illustration, where $\lambda_4^\downarrow$ is fixed by the normalization. Right: Two-dimensional projection of the three polytopes onto the hyperplane $\lambda_3=1$. The red vertex $\bd v=(2,1,1,0)$ marks the highest weight contained in all three polytopes. (See text for more explanations.) \label{fig:polytope_3D}}
\end{figure*}

By considering all valid permutations of natural occupation number vectors, we define in addition
\begin{align}
\Sigma^{(p)}_{S,M}&:= \left\{\bd\lambda^{(l)}\,\Big\vert\,\exists\gamma\in \mathcal{L}^1_{N,S,M}, \pi\in \mathcal{S}_{d}: \bd\lambda^{(l)} = \pi(\mathrm{spec}^\downarrow(\gamma))\right\},\label{eq:Sigma-p}\\
\Sigma^{(e)}_{S,M}&:= \left\{\bd\lambda^{(l)}\,\Big\vert\,\exists\gamma\in \xxbar{\mathcal{L}}^1_{N,S,M}, \pi\in \mathcal{S}_{d}: \bd\lambda^{(l)} = \pi(\mathrm{spec}^\downarrow(\gamma))\right\}\,,\label{eq:Sigma-e}
\end{align}
where $\mathcal{S}_d$ denotes the permutation group of degree $d$.
The set $\Sigma^{(e)}_{S,M}$ is described by the spin-adapted Pauli constraints derived in Ref.~\cite{LCLMS24} ($K \equiv (N-2S)/2$):
\begin{align}\label{eq:PC_SM}
& \hspace{-1.6cm} 2 \geq \lambda_1^{(l)} \geq \lambda_2^{(l)}  \geq \ldots \lambda_d^{(l)} \geq 0\,, \nonumber \\
\sum_{m=1}^{K+j}\lambda^{(l)}_m  &\leq N-2S +j\,,\quad   \forall j\in\{1,\ldots, 2S-1\} \,,\nonumber \\
\sum_{m=1}^d\lambda^{(l)}_m&=N\,.
\end{align}

Figure~\ref{fig:polytope_3D} illustrates the sets $\Sigma^{(e)}_{S,M}$ and $\Sigma^{(p)\downarrow}_{S,M}$ for $S=M=1$ and $N=d=4$, where the latter is described by the following spin-adapted GPCs \cite{Altunbulak-phd}
\begin{align}\label{eq:GPC_31}
\lambda^{(l)}_1 \geq \lambda^{(l)}_2 \geq \lambda^{(l)}_3 \geq \lambda^{(l)}_4 &\geq  0\,,\nonumber \\
\lambda^{(l)}_1 + \lambda^{(l)}_2+\lambda^{(l)}_3+\lambda^{(l)}_4 &= 4\,, \nonumber \\
\lambda^{(l)}_1 - \lambda^{(l)}_3 &\le 1\,, \nonumber  \\
\lambda^{(l)}_2 - \lambda^{(l)}_4 &\le 1 \,,  \nonumber  \\
2\lambda^{(l)}_1 - \lambda^{(l)}_2 + \lambda^{(l)}_4 &\le 3 \,,\nonumber  \\
\lambda^{(l)}_2 + 2\lambda^{(l)}_3 - \lambda^{(l)}_4 &\le 3 \,.
\end{align}
In the left panel, $\lambda_4^{(l)}$ is omitted since it is fixed by the normalization $\Tr[\gamma_l]=N$. For $S\neq 0$, the spin-adapted Pauli polytope $\Sigma^{(e)}_{S,M}$ (red) is more restrictive than the standard Pauli exclusion principle $0\leq \lambda_i^{(l)}\leq 2$ \cite{LCLMS24}. The polytope $\Sigma^{(p)\downarrow}_{S,M}$, characterized by spin-adapted GPCs, is shown in green. The right panel projects $\Sigma^{(p)\downarrow}_{S,M}$ onto the hyperplane $\lambda_3=1$. The vertex $\bd v = (2, 1,1,0)$ is the only shared vertex of $\Sigma^{(p)\downarrow}_{S,M}$ and $\Sigma^{(e)\downarrow}_{S,M}$, and it represents the analogue of the Hartree--Fock point $\bd\lambda_{\mathrm{HF}}=(1, \ldots, 1, 0, \ldots)$ in the space of full 1RDMs $\gamma$.

\subsection{Selection rules \label{sec:selection}}

In this section, we prove that if the natural orbital occupation number vector $\bd\lambda^{(l)}$ is pinned to the boundary of the spectral polytope $\Sigma^{(p)}_{S,M}$, the corresponding quantum state $\ket{\Psi} \in \mathcal{H}_N^{(S,M)}$ acquires a simplified structure. Furthermore, we demonstrate quantitatively that this structural simplification persists approximately in the case of quasipinning. In this way, we extend the key results \eqref{eq:DtoPsi} and \eqref{eq:Psi-selec} to explicitly incorporate spin degrees of freedom.

\subsubsection{Consequences of pinning}\label{sec:SSRPin}

Consider a pure state $\ket{\Psi} \in \mathcal{H}_N^{(S,M)}$ and a spin-adapted GPC,
\begin{equation}\label{eq:Dl}
D(\bl^{(l)})= \kappa_0 + \sum_{j=1}^{d} \kappa_j \lambda_j^{(l)} \geq 0\,,
\end{equation}
for the corresponding setting $(N,S,d)$. Analogous to Eq.~\eqref{eq:D-hat} for the spin-independent case, we define the associated operator
\begin{equation}\label{eq:Dhatl}
\hat{D}^{(\Psi)} \equiv \kappa_0 \mathbbm{1} + \sum_{j=1}^{d} \kappa_j \hat{n}_j\,,
\end{equation}
where $\hat{n}_j \equiv \hat{E}_{jj}$ is the particle number operator for the $j$-th natural orbital of $\ket{\Psi}$. This formulation allows us to interpret Eq.~\eqref{eq:Dl} as the expectation value of $\hat{D}^{(\Psi)}$,
\begin{equation}\label{eq:DlvsDlhat}
D(\bl^{(l)}) = \bra{\Psi} \hat{D}^{(\Psi)} \ket{\Psi}.
\end{equation}

To understand the implications of pinning, we seek a deformation of $\ket{\Psi}$ that reduces the corresponding value $D(\bl^{(l)})$. The hope is that this reveals structural constraints on $\ket{\Psi}$ when $\bl^{(l)}$ already saturates the GPC. We introduce the following deformation, evaluated for infinitesimal times $t\approx 0$:
\begin{eqnarray}\label{eq:flow0}
\ket{\Psi(t)} &:=& \frac{\mathrm{e}^{- t \hat{D}^{(\Psi)}}\ket{\Psi}}{\|\mathrm{e}^{- t \hat{D}^{(\Psi)}}\ket{\Psi}\|_2} \nonumber \\
&=& \left[1-t\,\big(\hat{D}^{(\Psi)}-D(\bl^{(l)})\big)\right] \ket{\Psi} + \mathcal{O}(t^2)\,,
\end{eqnarray}
where we used Eq.~\eqref{eq:DlvsDlhat} in the second line. Following a derivation analogous to the one in Appendix \ref{app:selection}, this deformation of $\ket{\Psi}$ changes as well the residual value $D(\bl^{(l)})$ according to
\begin{equation}
D(\bl^{(l)}(t)) =  D(\bl^{(l)})  - 2 \,t\, \text{Var}_{\Psi}\big(\hat{D}^{(\Psi)}\big)  + \mathcal{O}(t^2),
\end{equation}
where we introduce the variance
\begin{equation}
\text{Var}_{\Psi}\big(\hat{D}^{(\Psi)}\big) = \langle \big(\hat{D}^{(\Psi)} - \langle \hat{D}^{(\Psi)}\rangle_{\Psi}   \big)^2 \rangle_{\Psi}
\end{equation}
of $\hat{D}^{(\Psi)}$ in the state $\ket{\Psi}$.

If $\bl^{(l)}$ exactly saturates the spin-adapted GPC, $D(\bl^{(l)}) = 0$, then $\text{Var}_{\Psi}\big(\hat{D}^{(\Psi)}\big) = 0$, as any nonzero variance would imply a violation of $D \geq 0$ for small perturbations $t$. This means that $\ket{\Psi}$ must be an eigenstate of $\hat{D}^{(\Psi)}$, which, together with Eq.~\eqref{eq:DlvsDlhat}, implies
\begin{equation}\label{eq:PinDltoPsi}
D(\bl^{(l)}) = 0 \quad \Leftrightarrow \quad \hat{D}^{(\Psi)} \ket{\Psi} =0.
\end{equation}

To reformulate this key result of our work as a superselection rule, we expand $\ket{\Psi} \in \mathcal{H}_N^{(S,M)}$ according to \eqref{eq:PsiISM} using configuration state functions $\ket{I}$ built from its natural orbitals. Since the basis states $\ket{I}$ are constructed by applying negative root operators \eqref{eq:Eij} to the highest weight state \eqref{eq:Lambda-SM}, they are eigenstates of the orbital occupation number operators $\hat{n}_j$. Consequently, they are also eigenstates of the operator $\hat{D}^{(\Psi)}$ in Eq.~\eqref{eq:Dhatl}. Hence, relation \eqref{eq:PinDltoPsi} yields the following superselection rule
\begin{equation}\label{eq:Psi-selec-Dl}
D(\bl^{(l)})=0 \quad \Rightarrow \quad \ket{\Psi} = \sum_{I\in\mathcal{I}_{N,S,M}^{(D)}} c_I \ket{I}\,,
\end{equation}
where the index set $\mathcal{I}_{N,S,M}^{(D)}$ contains precisely those spin-configurations for which $\hat{D}^{(\Psi)}\ket{I}=0$.

\subsubsection{Consequences of quasipinning}\label{sec:SSRQPin}

In this section, we establish \emph{quantitatively} that Eq.~\eqref{eq:PinDltoPsi} and the superselection rule \eqref{eq:Psi-selec-Dl} remain approximately valid in the case of quasipinning, $D(\bl^{(l)}) \approx 0$, implying an approximate restriction on the wave function structure. To achieve this, we modify \eqref{eq:flow0} in such a way that it continuously transforms an initial state $\ket{\Psi}$ exhibiting quasipinning into one exhibiting pinning. This requires generalizing \eqref{eq:flow0} beyond infinitesimal times by introducing a time-dependent variant of $\hat{D}^{(\Psi)}$ that governs the imaginary time evolution. In analogy to the spin-independent derivation in Ref.~\cite{CBV17}, we define the following differential flow equation:
\begin{equation}\label{eq:flow}
\frac{\mathrm{d}}{\mathrm{d}t}\ket{\Psi(t)}= -\big(\mathbbm{1}-\ket{\Psi(t)}\!\bra{\Psi(t)}\big)\,\hat{D}^{(\Psi(t))} \,\ket{\Psi(t)}\,,
\end{equation}
where the projector $\mathbbm{1}-\ket{\Psi(t)}\!\bra{\Psi(t)}$ ensures proper normalization of $\ket{\Psi(t)}$ at all times.

In Appendix \ref{app:selection}, we show that the evolution of the corresponding $D(\bd\lambda^{(l)}(t))$ satisfies
\begin{equation}\label{eq:Delta-variance}
\frac{\mathrm{d}}{\mathrm{d}t}D(\bd\lambda^{(l)}(t)) = - 2\,\text{Var}_{\Psi(t)}\big(\hat{D}^{(\Psi(t))}\big),
\end{equation}
where $\hat{D}^{(\Psi(t))}$ is given by Eq.~\eqref{eq:Dhatl} with $\ket{\Psi} \equiv \ket{\Psi(t)}$. Since the variance is always non-negative, Eq.~\eqref{eq:Delta-variance} confirms that the time evolution in \eqref{eq:flow} continuously reduces $D(\bd\lambda^{(l)}(t))$, meaning that the vector $\bd\lambda^{(l)}(t)$ approaches the facet of the polytope $\Sigma_{S,M}^{(p)}$ corresponding to pinning by the GPC $D\geq 0$.

Assuming that initially $D(\bd\lambda^{(l)}(0))\ll 1$, it follows analogously to the derivation presented in Ref.~\cite{CBV17} for the spin-independent case, that
\begin{equation}\label{eq:bound-Psi}
\big\| \ket{\Psi(t\to \infty)}- \ket{\Psi(0)}\big\|_2 \leq  \sqrt{2 D \big(\bd\lambda^{(l)}(0)\big)},
\end{equation}
and $D \big(\bd\lambda^{(l)}(t\to \infty)\big)=0$, where $\bd\lambda^{(l)}(0)$ is the natural orbital occupation number vector of the initial state $\ket{\Psi(0)} \equiv \ket{\Psi}$. Since the evolved state $\ket{\Psi(t\to \infty)}$ exhibits pinning by the GPC $D\geq 0$, the estimate \eqref{eq:bound-Psi} implies that the original state $\ket{\Psi}$ approximately exhibits the structure associated with pinning. The deviation of $\ket{\Psi}$ from an exactly pinned state, measured in the $L^2$-norm, is bounded above by $\sqrt{2 D \big(\bd\lambda^{(l)}(0)\big)}$. This confirms quantitatively the stability of the key results \eqref{eq:PinDltoPsi} and \eqref{eq:Psi-selec-Dl} in case of quasipinning.

\subsection{Vertices versus configuration state functions}

\begin{figure}[tb]
\centering
\includegraphics[width=0.75\linewidth]{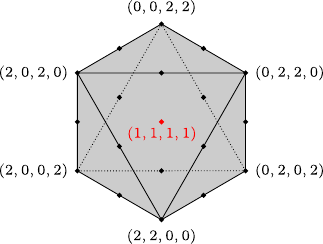}
\caption{Spectral polytope $\Sigma_{0,0}^{(e)}=\Sigma_{0,0}^{(p)}$ and weight lattice for $N=d=4, S=0$. Configuration states whose weights, i.e., natural orbital occupation number vectors $\bl^{(l)}$, are not permutations of the Hartree--Fock point $\bd v=(2,2,0,0)$ map to non-extremal points on the edges (permutations of $(2,1,1,0)$) or the center $(1,1,1,1)$ of $\Sigma_{0,0}^{(e)}$. \label{fig:weight-lattice}}
\end{figure}
Inspired by the superselection rule derived in Sec.~\ref{sec:selection}, we further investigate the relationship between the vertices of the spectral polytopes and the configuration states $\ket{I}$ used to expand $\ket{\Psi} \in \mathcal{H}_N^{(S,M)}$. In particular, we emphasize the potential differences from the spin-independent case. Notably, two distinctive features arise in the spin-dependent setting that do not occur in the spin-independent case.

First, in the spin-independent case, configuration states are simply Slater determinants. In contrast, spin-adapted configuration states $\ket{I}, I \in \mathcal{I}_{N,S,M}$ can be linear combinations of Slater determinants. Consequently, some configuration states $\ket{I}$ do not correspond to vertices of the spectral polytopes $\Sigma_{S,M}^{(p)}$ and $\Sigma_{S,M}^{(e)}$. Following representation theory terminology (see \cite{Hall15} for further details), we refer to the natural orbital occupation number vectors $\bd \lambda^{(l)}$ associated with configurations $I \in \mathcal{I}_{N,S,M}$ as \emph{weights}. The collection of these weights forms the \emph{weight lattice}, whose convex hull defines the spectral set $\Sigma_{S,M}^{(e)}$.

Figure~\ref{fig:weight-lattice} illustrates $\Sigma_{0,0}^{(e)}$ and its weight lattice for $N=d=4$ and $S=0$. The Hilbert space $\mathcal{H}_4^{(0,0)}$ is 20-dimensional. Since no spin-adapted GPCs exist for $S=0$ beyond the Pauli exclusion principle, $0 \leq \lambda_i^{(l)} \leq 2$, the spectral sets coincide, $\Sigma_{0,0}^{(e)}=\Sigma_{0,0}^{(p)}$. Notably, only the six vertices of $\Sigma_{0,0}^{(e)}$ corresponding to permutations of the Hartree--Fock point $\bd v = (2,2,0,0)$ represent Slater determinant states. Additionally, twelve configuration states map to points along the edges of $\Sigma_{0,0}^{(e)}$, while two configuration states correspond to the center point $\bd \lambda^{(l)}=(1,1,1,1)$ (marked in red).

A second fundamental difference from the spin-independent case is that in the context of the self-consistent expansion \eqref{eq:PsiISM} a nonzero expansion coefficient $c_I$ for configuration states corresponding to the interior of $\Sigma_{S,M}^{(p)}$ precludes pinning for \emph{all} spin-adapted GPCs. Specifically, in Fig.~\ref{fig:weight-lattice}, if either of the configuration states associated with $(1,1,1,1)$ contributes to the wave function expansion \eqref{eq:PsiISM}, then, by virtue of the superselection rule \eqref{eq:Psi-selec-Dl}, pinning is impossible since none of the generalized spin-adapted Pauli constraints can be exactly saturated.

\section{Pinning in the Borland--Dennis setting\label{sec:BD}}

\begin{figure}[tb]
\centering
\includegraphics[width=\linewidth]{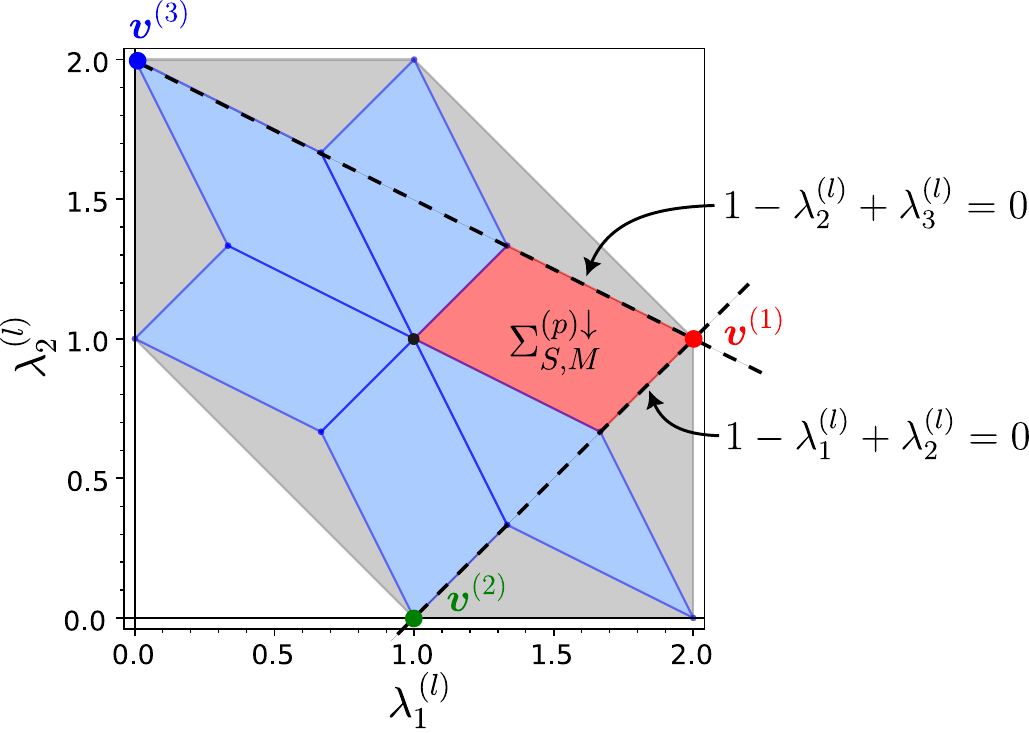}
\caption{Illustration of the spectral polytope $\Sigma^{(p)\downarrow}_{S,M}$ and the spin-adapted GPCs for the Borland--Dennis setting with $N=3$, $2d=6$ and $S=M=1/2$.  \label{fig:BD_2d}}
\end{figure}

A key question emerges from the previous sections: Is the (quasi)pinning observed in Refs.~\cite{SGC13, BGS13, Schilling15, S15-H, CM15, TEVS16, TVS16, CBV17, LS18, SAKL18, SBLMS20} at the level of the full 1RDM reflecting a hidden structural feature of the wave function, or is it merely a consequence of spin symmetries? To elaborate on this issue, we examine the Borland--Dennis setting \cite{BD72, R07} with $N=d=3$.

The solution to the spin-independent variant of this problem was originally derived by Borland and Dennis in 1972 \cite{BD72} and later rigorously proven in Ref.~\cite{R07}. The corresponding convex polytope of admissible natural occupation numbers $\bd\lambda = \bd\lambda^\downarrow$ is constrained by the following conditions \cite{BD72, R07}:
\begin{align}\label{eq:BD-cons}
&\lambda_1\geq \lambda_2\geq \lambda_3\geq \lambda_4\geq \lambda_6 \geq 0,\nonumber\\
&\lambda_1+\lambda_6 = \lambda_2+\lambda_5=\lambda_3+\lambda_4=1, \nonumber\\
&D(\bd\lambda)=\lambda_5+\lambda_6-\lambda_4\geq 0.
\end{align}

For $S=M=1/2$, any pure state $\ket{\Psi}\in\mathcal{H}_3^{(1/2,1/2)}$ can be expressed as \cite{TLH18, LMS24}
\begin{equation}\label{eq:Psi-BD}
|\Psi\rangle = c_{12\tilde{1}}|1\!\uparrow, 2\!\uparrow, \tilde{1}\!\downarrow\rangle+ c_{13\tilde{2}}|1\!\uparrow, 3\!\uparrow, \tilde{2}\!\downarrow\rangle+ c_{23\tilde{3}}|2\!\uparrow, 3\!\uparrow, \tilde{3}\!\downarrow\rangle,
\end{equation}
where $|c_{12\tilde{1}}| \geq |c_{13\tilde{2}}| \geq |c_{23\tilde{3}}|$, and $\{\ket{j}\}_{j=1}^3$, $\{\ket{\tilde{j}}\}_{j=1}^3$ form two orthonormal sets of orbitals. As demonstrated in Refs.~\cite{TLH18, LMS24}, the spatial components of the natural spin-orbitals must additionally satisfy
\begin{equation}\label{eq:BD-NO}
\mathrm{e}^{\mathrm{i}\varphi_1}\sqrt{\lambda_{\tilde{1}\downarrow}}\langle3|\tilde{1}\rangle-\mathrm{e}^{\mathrm{i}\varphi_2}\sqrt{\lambda_{\tilde{2}\downarrow}}\langle 2|\tilde{2}\rangle+\sqrt{\lambda_{\tilde{3}\downarrow}}\langle1|\tilde{3}\rangle=0,
\end{equation}
where $\varphi_i\in\RR$ for $i\in\{1,2,3\}$. This condition highlights that the solution to the pure one-body $(N,S,M)$-representability problem is inherently dependent not only on the natural occupation numbers but also on the natural spin-orbitals, as discussed in Sec.~\ref{sec:orbital}.

The 1RDM $\gamma$ of the state $|\Psi\rangle$ is diagonal in the natural spin-orbital basis $\{\ket{i\!\uparrow}, \ket{\tilde{i}\!\downarrow}\}$. Consequently, the three equalities in Eq.~\eqref{eq:BD-cons} directly translate into
\begin{equation}\label{eq:BD-equalities}
\lambda_{1\uparrow} +\lambda_{\tilde{3}\downarrow}=\lambda_{2\uparrow} +\lambda_{\tilde{2}\downarrow} = \lambda_{3\uparrow} +\lambda_{\tilde{1}\downarrow}=1.
\end{equation}
Furthermore, the structure of the quantum state $\ket{\Psi}$ in Eq.~\eqref{eq:Psi-BD} results in the following ordering relations:
\begin{align}
\lambda_{1\uparrow}&\geq \lambda_{\tilde{1}\downarrow}, \quad \lambda_{1\uparrow}\geq\lambda_{\tilde{2}\downarrow},\nonumber\\
\lambda_{2\uparrow}&\geq \lambda_{\tilde{1}\downarrow}, \quad \lambda_{2\uparrow}\geq \lambda_{\tilde{3}\downarrow},\nonumber\\
\lambda_{3\uparrow}&\geq \lambda_{\tilde{2}\downarrow}, \quad \lambda_{3\uparrow}\geq \lambda_{\tilde{3}\downarrow}.
\end{align}
Thus, only two possible orderings of the six natural occupation numbers are possible:
\begin{align}
\text{(i)}&\quad \lambda_{1\uparrow}\geq \lambda_{2\uparrow}\geq \lambda_{3\uparrow}\geq \lambda_{\tilde{1}\downarrow}\geq \lambda_{\tilde{2}\downarrow}\geq \lambda_{\tilde{3}\downarrow},\nonumber\\
\text{(ii)}&\quad \lambda_{1\uparrow}\geq \lambda_{2\uparrow}\geq \lambda_{\tilde{1}\downarrow}\geq \lambda_{3\uparrow}\geq \lambda_{\tilde{2}\downarrow}\geq \lambda_{\tilde{3}\downarrow}.
\end{align}
The first one corresponds to $|c_{13\tilde{2}}|^2+ |c_{23\tilde{3}}|^2  \geq |c_{12\tilde{1}}|^2$ and the second one to $|c_{13\tilde{2}}|^2+ |c_{23\tilde{3}}|^2  \leq |c_{12\tilde{1}}|^2$.
For non-degenerate natural occupation numbers, ordering (i) never results in pinning of the GPC in Eq.~\eqref{eq:BD-cons},
\begin{equation}
D(\bl) = |c_{13\tilde{2}}|^2+ |c_{23\tilde{3}}|^2 - |c_{12\tilde{1}}|^2 > 0 \,,
\end{equation}
whereas ordering (ii) inherently implies pinning,
\begin{equation}
D(\bl) = |c_{13\tilde{2}}|^2+ |c_{23\tilde{3}}|^2 - (|c_{13\tilde{2}}|^2+ |c_{23\tilde{3}}|^2) = 0\,.
\end{equation}

Furthermore, in case (ii), we have $\hat{D}^{(\Psi)} = \hat{n}_{\tilde{2}\downarrow}+\hat{n}_{\tilde{3}\downarrow} - \hat{n}_{3\uparrow}$. The selection rule $\hat{D}^{(\Psi)}\ket{\Psi}=0$, as discussed in Secs.~\ref{sec:spin-symm} and \ref{sec:selection}, dictates that the state $\ket{\Psi}$ takes precisely the form given in Eq.~\eqref{eq:Psi-BD}. Since \emph{any} pure state in the spin sector $\mathcal{H}_{3}^{(1/2,1/2)}$ can be expressed in this form, pinning does not provide any additional structural information beyond confirming that $|c_{12\tilde{1}}|^2 \geq |c_{13\tilde{2}}|^2+ |c_{23\tilde{3}}|^2 $, corresponding to case (ii).

In contrast, pinning at the level of the orbital 1RDM leads to a further simplification of the structure \eqref{eq:Psi-BD}, as demonstrated in the following.
The spectral polytope $\Sigma_{S,M}^{(p)\downarrow}$ for the Borland--Dennis setting in the spin sector $S=M=1/2$ is described by two spin-adapted GPCs \cite{AK08}:
\begin{align}\label{eq:BD-GPC-spin}
D_1(\bd\lambda^{(l)}) &= 1 - \lambda_1^{(l)} + \lambda_2^{(l)}\geq 0\,,\nonumber\\
D_2(\bd\lambda^{(l)}) &= 1 - \lambda_2^{(l)} + \lambda_3^{(l)} \geq 0\,.
\end{align}
We illustrate this polytope in Fig.~\ref{fig:BD_2d} in red. The non-convex set $\Sigma_{S,M}^{(p)}$, which includes all $3!=6$ orderings of $\lambda_1^{(l)},\lambda_2^{(l)},\lambda_3^{(l)}$, is also shown (blue and red). Additionally, the three vertices $\bd v^{(i)}$, $i=1, 2, 3$, marked by the red, green, and blue dots in Fig.~\ref{fig:BD_2d}, correspond to the configuration states:
\begin{align}
\ket{I_1}&\equiv \ket{1\!\uparrow,  1\!\downarrow, 2\!\uparrow}\,,\nonumber\\
\ket{I_2}&\equiv E_{31}^\dagger E_{32}^\dagger \ket{I_1} =  \ket{1\!\uparrow, 3\!\uparrow, 3\!\downarrow }\,,\nonumber\\
\ket{I_3}&\equiv E_{31}^\dagger E_{21}^\dagger \ket{I_1} = \ket{2\!\uparrow , 2\!\downarrow, 3\!\uparrow}\,,
\end{align}
where $\ket{I_1}$ is the highest weight state from Eq.~\eqref{eq:Lambda-hwv}. Additional configuration states, defined as $\ket{I_4} \equiv  \ket{1\!\uparrow, 2\!\uparrow, 2\!\downarrow }, \ket{I_5} \equiv  \ket{2\!\uparrow, 3\!\uparrow, 3\!\downarrow }, \ket{I_6} \equiv  \ket{1\!\uparrow, 1\!\downarrow, 3\!\uparrow }$, are obtained via successive applications of negative root operators $E_{j+1, j}^\dagger$. Their occupation number vectors yield the remaining three vertices of the gray polytope $\Sigma_{S,M}^{(e)}$. Additionally, two configuration states, denoted $\ket{\tilde{I}_{7/8}}$, map to the weight $(1,1,1)$. These states are generated from the highest weight state $\ket{I_1}\equiv\ket{1\!\uparrow,1\!\downarrow, 2\uparrow}$ via the operators $E_{21}^\dagger E_{32}^\dagger$ and $E_{32}^\dagger E_{21}^\dagger$, leading to (after normalization) \footnote{In general, if a configuration state maps to a weight non-uniquely, the resulting states from monomials of negative fundamental root operators $E_{j+1, j}^\dagger$ are typically linearly independent but not orthonormal.}:
\begin{align}
\ket{\tilde{I}_{7}} &= \frac{1}{\sqrt{2}}\left(\ket{1\!\uparrow, 2\!\uparrow, 3\!\downarrow} - \ket{1\!\uparrow, 2\!\downarrow, 3\!\uparrow}\right)\,,\nonumber\\
\ket{\tilde{I}_{8}} &= \frac{1}{\sqrt{2}}\left(\ket{1\!\uparrow, 2\!\downarrow, 3\!\uparrow} - \ket{1\!\downarrow, 2\!\uparrow, 3\!\uparrow}\right)\,.
\end{align}
These states can be further orthogonalized to yield:
\begin{align}
\ket{I_{7/8}} &= \frac{1}{\sqrt{3}}\left(\ket{1\!\uparrow, 2\!\uparrow, 3\!\downarrow}+\mathrm{e}^{\pm \mathrm{i} 2\pi/3}\ket{1\!\uparrow, 2\!\downarrow, 3\!\uparrow}\right.\nonumber\\
&\quad \left.+\mathrm{e}^{\mp \mathrm{i} 2\pi/3}\ket{1\!\downarrow, 2\!\uparrow, 3\!\uparrow}\right)\,.
\end{align}

Any state $\ket{\Psi}\in\mathcal{H}_3^{(1/2,1/2)}$ can then be expanded as
\begin{equation}\label{eq:Psi_BD_I}
\ket{\Psi} = \sum_{i=1}^8c_{i}\ket{I_i}\,.
\end{equation}
If, in addition, the orbitals entering the eight spin-adapted configuration states $\ket{I_i}$ are the natural orbitals of $\ket{\Psi}$, this leads to additional quadratic conditions on the expansion coefficients $c_i$ which further restrict the expansion in Eq.~\eqref{eq:Psi_BD_I}.  

To analyze the impact of pinning by spin-adapted GPCs, we assume $D_1(\bd\lambda^{(l)})=0$. According to the superselection rule \eqref{eq:Psi-selec-Dl}, this implies that any state $\ket{\Psi} \in \mathcal{H}_3^{(1/2,1/2)}$ with natural orbital occupation numbers $\bd\lambda^{(l)}$ can be expressed as a linear combination of the configuration states $\ket{I_1}$ and $\ket{I_2}$. Geometrically, this means that only configuration states whose occupation number vectors lie on the hyperplane $D_1\equiv 0$ contribute --- here those mapping to the vertices $\bd v^{(1)} = (2,1,0)$ (red) and $\bd v^{(2)}=(1,0,2)$ (green). Similarly, if $D_2(\bd\lambda^{(l)})=0$, the wave function simplifies to $\ket{\Psi} = c_1 \ket{I_1} + c_3\ket{I_3}$ with $c_1, c_3\in \CC$ and $|c_1|^2+ |c_3|^2=1$. In the case where both GPCs are saturated, $D_1(\bd\lambda^{(l)})=D_2(\bd\lambda^{(l)})=0$, we have $\ket{\Psi}=\ket{I_1}$.

This analysis confirms that pinning by spin-independent GPCs can arise purely due to spin symmetries of the $N$-electron quantum state $\ket{\Psi}$ and thus should be considered trivial, as it does not further simplify the wave function expansion. In contrast, pinning by spin-adapted GPCs leads to an actual structural restriction. Moreover, these conclusions extend naturally to quasipinning in virtue of the estimate \eqref{eq:bound-Psi}.

Finally, we illustrate these findings through an exemplary state exhibiting quasipinning. This example reveals a subtle but crucial aspect: (Quasi)pinning can be trivial not only due to symmetry constraints but also as a direct consequence of weak correlation, or more generally, due to (quasi)pinning by (spin-adapted) Pauli constraints. To illustrate this, we consider the state
\begin{equation}\label{eq:BD-Psi-ex}
\ket{\Psi} = \sqrt{\frac{2}{3}-\varepsilon^2} \ket{I_1} +\sqrt{\frac{1}{3}} \ket{I_3}+ \varepsilon\ket{I_2}+\mathcal{O}(\varepsilon^2)\,,
\end{equation}
where $\varepsilon \ll 1$ and in leading order $(\varepsilon^2)$ neither $\ket{I_1}$ nor $\ket{I_3}$ contribute to the generic terms $\mathcal{O}(\varepsilon^2)$. Then, the spin-independent GPC in Eq.~\eqref{eq:BD-cons} satisfies
\begin{equation}
D(\bd\lambda) = \lambda_5+\lambda_6-\lambda_4 = \mathcal{O}(\varepsilon^3)\,,
\end{equation}
indicating quasipinning with leading order proportional to $\varepsilon^3$, while for the spin-adapted GPCs \eqref{eq:BD-GPC-spin} we obtain
\begin{equation}\label{eq:d1-example}
D_1(\bd\lambda^{(l)}) = 1+\mathcal{O}(\varepsilon^3)\,,\quad D_2(\bd\lambda^{(l)}) = 3 \varepsilon^2 +\mathcal{O}(\varepsilon^3)\,.
\end{equation}
Thus, in absolute terms, $D_2(\bd\lambda^{(l)})$ is one order of magnitude larger than $D(\bd\lambda)$, reflecting that one order of quasipinning by the spin-independent GPC originates from spin symmetries.

To determine whether quasipinning is non-trivial in the sense that it is enforced by quasipinning of (spin-adapted) Pauli constraints (see also Sec.~\ref{sec:selection}), we analyze its relation to the Pauli constraints. In the spin-independent case, the distance of $\bl$ to the Hartree--Fock point $\bd v_{\mathrm{HF}}=(1, \ldots, 1, 0,  \ldots)$ is given by
\begin{equation}\label{eq:BD-X1}
X_1(\bd\lambda) \equiv 3-\lambda_1-\lambda_2-\lambda_3 = \frac{2}{3}+2\varepsilon^2+\mathcal{O}(\varepsilon^3)\,.
\end{equation}
Thus, the ratio $D/X_{1}$ remains of order $\varepsilon^3$, implying that the quasipinning of order $\varepsilon^3$ cannot (even not partially) be attributed to weak correlation. However, since $\lambda_1=1$ enforces pinning, as confirmed by the estimate
\begin{eqnarray}
D(\bl) &=& 2 -\lambda_1 -\lambda_2-\lambda_4 \nonumber \\
&=& 1 -\lambda_1 -\lambda_2+\lambda_3 \leq 1-\lambda_1\,,
\end{eqnarray}
we also compute
\begin{align}\label{eq:BD-X2}
X_2(\bd\lambda) \equiv 1-\lambda_1 = \varepsilon^2+\mathcal{O}(\varepsilon^3)\,.
\end{align}
Therefore, comparing $D$ and $X_{2}$ confirms that two orders of quasipinning can be attributed to quasipinning by the Pauli constraint $X_{2}\geq 0$.

Similarly, in the spin-adapted setting, we find that both $D_{1/2}(\bd\lambda^{(l)})$ intersect with the hyperplanes of the spin-adapted Pauli constraints $X^{(l)}_{1/2}$, which satisfy
\begin{align}
X_1^{(l)}(\bd\lambda^{(l)}) &\equiv 2-\lambda_1^{(l)} =\frac{2}{3}+\varepsilon^2+\mathcal{O}(\varepsilon^3)\,,\nonumber\\
X_2^{(l)}(\bd\lambda^{(l)}) &\equiv 3-\lambda_1^{(l)}-\lambda_2^{(l)} = \frac{1}{3}+\mathcal{O}(\varepsilon^3)\,.
\end{align}
Thus, the ratio between $D_2(\bd\lambda^{(l)})$ and $X_i^{(l)}$ is of order $\varepsilon^2$. This confirms that the quasipinning of $D_2(\bd \lambda^{(l)})$ is one order of magnitude less trivial than that of $D(\bd\lambda)$ when compared to the Pauli constraints.

\section{Spin-adapted $Q$-parameter\label{sec:Q-param}}

As motivated in the previous section, we introduce a \emph{spin-adapted} $Q$-parameter, extending the $Q$-parameter from the spin-independent case \cite{TVS17} to the more general, and practically relevant, spin-dependent setting. The $Q$-parameter serves as a crucial tool in complementing the (quasi)pinning analysis, as it systematically quantifies the extent to which (quasi)pinning of spin-adapted GPCs is induced by ordinary or spin-adapted Pauli constraints.

The construction of the $Q$-parameter is based on the inclusion relation
\begin{equation}\label{eq:Pinclusion}
\Sigma_{S,M}^{(p)\downarrow} \subseteq \Sigma_{S,M}^{(e)} \subseteq \Sigma^{(e)}\,.
\end{equation}
This hierarchy follows directly from the definitions of the three polytopes $\Sigma_{S,M}^{(p) \downarrow}, \Sigma_{S,M}^{(e)}$, and $\Sigma^{(e)}$, which are the spectral sets of orbital 1RDMs $\gamma_l$ corresponding to three different sets of $N$-fermion states. The latter satisfy an analogous inclusion relation,
\begin{equation}
\mathcal{P}^N_{S,M} \subseteq \mathcal{E}^N_{S,M} \subseteq \mathcal{E}^N\,,
\end{equation}
where $\mathcal{P}^N_{S,M}$ represents pure states in the $(S,M)$-spin sector $\mathcal{H}_N^{(S,M)}$, $\mathcal{E}^N_{S,M}$ denotes the set of all ensemble states on $\mathcal{H}_N^{(S,M)}$, and $\mathcal{E}^N$ is the space of general ensemble $N$-fermion states on $\mathcal{H}_N$.

The inclusion relation \eqref{eq:Pinclusion} is illustrated in Fig.~\ref{fig:polytope_3D} for an exemplary setting. From this visualization, it becomes evident that the three sets $\Sigma_{S,M}^{(p) \downarrow}, \Sigma_{S,M}^{(e)}, \Sigma^{(e)}$ share common boundaries. Consequently, if a point $\bl \in \Sigma_{S,M}^{(p)\downarrow}$ is close to the boundary of $\Sigma_{S,M}^{(e)}$ or $\Sigma^{(e)}$, then it must necessarily be at least as close to the boundary of $\Sigma_{S,M}^{(p)\,\downarrow}$. This observation implies that (quasi)pinning by (spin-adapted) Pauli constraints lead to apparent (quasi)pinning of spin-adapted GPCs.

To formalize this point, we establish a relation between spin-adapted GPCs and spin-adapted Pauli constraints by identifying linear dependencies among them. Specifically, for each spin-adapted GPC $D\geq 0$, we determine the minimal set of (spin-adapted) Pauli constraints whose simultaneous saturation would necessarily imply pinning of the GPC. Geometrically, this corresponds to identifying, for each non-trivial facet of $\Sigma_{S,M}^{(p)\downarrow}$ (described by $D\equiv 0$), the largest possible face of the surrounding polytope --- either $\Sigma_{S,M}^{(e)}$ or $\Sigma^{(e)}$ --- that contains this facet.

Based on this construction, we define the spin-adapted $Q$-parameter. Since the polytope $\Sigma_{S,M}^{(p)\downarrow}$ is enclosed within two distinct reference polytopes in \eqref{eq:Pinclusion}, this procedure yields two alternative formulations of the spin-adapted $Q$-parameter, depending on the choice of the surrounding polytope:
\begin{enumerate}
    \item \textbf{Case A:} $\Sigma^{(e)}_{S,M}$, characterized by spin-adapted Pauli constraints \eqref{eq:PC_SM}.
    \item \textbf{Case B:} $\Sigma^{(e)}$, characterized by the ordinary Pauli constraints, thus disregarding additional spin-adapted Pauli constraints \eqref{eq:PC_SM}.
\end{enumerate}

In Fig.~\ref{fig:polytope_3D}, Case A corresponds to the red polytope, while Case B corresponds to the Pauli polytope outlined by gray dashed lines. Case A is conceptually more relevant, as it ensures $(N,S,M)$-representability at both the GPC and Pauli constraint levels. However, including Case B allows for a direct comparison with previous results from Refs.~\cite{TVS17, SAKL18}.

Since $\Sigma^{(e)}_{S,M}$ exhibits a more intricate geometric structure than the Pauli polytope $\Sigma^{(e)}$ (see Ref.~\cite{LMS24}), deriving the spin-adapted $Q$-parameter requires refining the methods from Ref.~\cite{TVS17} rather than directly applying them. Moreover, $\Sigma^{(e)}$ is a simplex only for even $N$, but not for odd $N$. In the latter case, $\Sigma^{(e)}$ corresponds to the permutohedron of the vertex $\bd v = (2, ..., 2, 1, 0, ..., 0)$.

We proceed by first deriving the symmetry-adapted $Q$-parameter for Case A and subsequently discussing Case B. Before doing so, we clarify some essential terminology: The dimension of a convex polytope is defined as the dimension of its affine hull. Zero-dimensional faces are called \emph{vertices}, one-dimensional faces are \emph{edges}, and faces of co-dimension 1, i.e., those of maximal dimension, are referred to as \emph{facets}. This distinction between general faces and facets is crucial and will be used throughout the discussion.

We use calligraphic notation, $\mathcal{F}$, to denote facets, while non-calligraphic notation, $F$, refers to general $k$-dimensional faces. The facet of $\Sigma_{S,M}^{(p)\downarrow}$ associated with the pinning of the $i$-th GPC, $D_i(\bd\lambda^{(l)})=0$, is denoted by $\mathcal{F}_i^{(p)}$. Similarly, facets of the convex polytope $\Sigma_{S,M}^{(e)\downarrow}$ are denoted by $\mathcal{F}_x^{(e)}$, where $x$ indexes different facets.

To compute the symmetry-adapted $Q$-parameter for a given spin-adapted GPC $D_i\geq 0$, we determine the intersections between its corresponding facet $\mathcal{F}_i^{(p)}$ and the boundary $\partial\Sigma_{S,M}^{(e)\downarrow}$ of $\Sigma_{S,M}^{(e)\downarrow}$. The resulting set of intersection elements is denoted by
\begin{equation}\label{eq:Ki}
\mathcal{K}_i:= \left\{M_x^{(i)}\,|\,M_x^{(i)}\equiv \mathcal{F}_x^{(e)}\cap \mathcal{F}_i^{(p)}\text{ for some }\mathcal{F}_x^{(e)} \right\}.
\end{equation}
Elements $M_x^{(i)}\in \mathcal{K}_i$ are at most $(d-2)$-dimensional. Moreover, the set $\mathcal{K}_i$ exhibits a natural hierarchical structure in which some of its elements are related by set inclusion (see Ref.~\cite{Ziegler-polytopes}), i.e., $M_x^{(i)} \subseteq M_{x^\prime}^{(i)}$. This ordering plays a crucial role in the computation of the spin-adapted $Q$-parameter (see also Ref.~\cite{TVS17}).

In fact, elements $M_x^{(i)}$ satisfying $M_x^{(i)}\subseteq M_{x^\prime}^{(i)}$ for some $M_{x^\prime}^{(i)}\in \mathcal{K}_i$ are redundant, as they obey the relation
\begin{equation}
\mathrm{dist}(\bl^{(l)},\mathcal{F}_i^{(p)}) \leq \mathrm{dist}(\bl^{(l)},M_{x^\prime}^{(i)}) \leq  \mathrm{dist}(\bl^{(l)},M_{x}^{(i)})\,.
\end{equation}
An instructive example of this was provided at the end of Sec.~\ref{sec:BD}, for the explicit quantum state \eqref{eq:BD-Psi-ex} (recall also Eqs.~\eqref{eq:BD-X1} and \eqref{eq:BD-X2}). There, in the spin-independent setting, we compared the distances of $\bl$ to the Hartree--Fock point (a zero-dimensional face, $M_1$) and to the facet $M_2$ given by $\lambda_1 \equiv 1$. The distance to $M_2$ is necessarily smaller than that to $M_1$. Consequently, a conclusive analysis of the non-triviality of quasipinning by the GPC in \eqref{eq:BD-equalities} requires comparing $D(\bl)$ to $X_2(\bl)$, while the comparison to $X_1(\bl)$ is redundant.

Following these geometric insights, it suffices to retain only those elements $M_x^{(i)}\in \mathcal{K}_i$ that are not contained in any other element of $\mathcal{K}_i$. Formally, this means computing the set
\begin{equation}\label{eq:Ii}
\mathcal{J}_i := \left\{M_x^{(i)}\in \mathcal{K}_i\,|\,\forall M_y^{(i)}\in \mathcal{K}_i:  M_x^{(i)} \subseteq  M_y^{(i)} \Rightarrow x=y\right\}.
\end{equation}
This ensures that each element in $\mathcal{J}_i$ represents a maximal independent intersection relevant for defining the spin-adapted $Q$-parameter.

Since each $M_x^{(i)}$ is contained within $\mathcal{F}_i^{(p)}$, the $l^1$-distance of an orbital natural occupation number vector $\bd\lambda^{(l)}\in\Sigma^{(p)\downarrow}_{S,M}$ is bounded from above as follows:
\begin{equation}\label{eq:ineq}
\mathrm{dist}_1\left(\bd\lambda^{(l)}, \mathcal{F}_i^{(p)}\right)\leq c_x^{(i)} \mathrm{dist}_1\left(\bd\lambda^{(l)}, M_x^{(i)}\right)\quad \forall M_x^{(i)}\in \mathcal{J}_i\,.
\end{equation}
The optimal constant $c_x^{(i)}\leq 1$ in Eq.~\eqref{eq:ineq} is given by
\begin{equation}
c_x^{(i)} := \max_{\bd\lambda\in\Sigma_{S,M}^{(p)\downarrow}}\left(\frac{\mathrm{dist}_1\left(\bd\lambda^{(l)}, \mathcal{F}_i^{(p)}\right)}{\mathrm{dist}_1\left(\bd\lambda^{(l)}, M_x^{(i)}\right)} \right)\,.
\end{equation}
This naturally leads, in analogy to Ref.~\cite{TVS17}, to a face $F_x^{(e)}$-induced spin-adapted auxiliary parameter,
\begin{equation}
Q_x^{(i)}(\bd\lambda^{(l)}):= -\log_{10}\left(\frac{\mathrm{dist}_1\left(\bd\lambda^{(l)}, \mathcal{F}_i^{(p)}\right)}{c_x^{(i)}\mathrm{dist}_1\left(\bd\lambda^{(l)}, M_x^{(i)}\right)} \right)\,.
\end{equation}
Maximizing this over all $M_x^{(i)}\in \mathcal{J}_i$,
\begin{equation}\label{eq:Qi}
Q^{(i)}(\bd\lambda^{(l)}):= \max_{M_x^{(i)}\in \mathcal{J}_i}Q_x^{(i)}(\bd\lambda^{(l)})\,,
\end{equation}
yields the sought-after spin-adapted $Q$-parameter for the GPC $D_i \geq 0$. The definitions in Eqs.~\eqref{eq:ineq}-\eqref{eq:Qi} can be extended analogously to any other $l^p$-distance metric.

Furthermore, a global $Q$-parameter, incorporating all spin-adapted GPCs, is defined as
\begin{equation}\label{eq:Q-global}
Q (\bd\lambda^{(l)}):= \max_{i} Q^{(i)}(\bd\lambda^{(l)})\,.
\end{equation}
Accordingly, the $Q$-parameter quantifies on a logarithmic scale the extent to which quasipinning by spin-adapted GPCs arises from quasipinning by spin-adapted Pauli constraints. For example, a value of $Q(\bl^{(l)})=2$ indicates that $\bl^{(l)}$ is $10^2=100$ times closer to the boundary of $\Sigma_{S,M}^{(p)}$ than expected from its proximity to the boundary of $\Sigma_{S,M}^{(e)}$.

The derivation for Case B) follows analogously, replacing $\Sigma^{(e)}_{S,M}$ with $\Sigma^{(e)}$.

For both settings A) and B), we explicitly determine the set $\mathcal{J}_i$ and the constants $c_x^{(i)}$ required to compute the spin-adapted $Q$-parameter $Q^{(i)}$ for various spin-adapted GPCs $D_i$. This encompasses all settings $(N,d,S,M)$ solved in Ref.~\cite{AK08}, as well as additional ones solved in this work. To achieve this, we analyze the $l^1$-norm and determine the sets $\mathcal{K}_i, \mathcal{J}_i$ as defined in Eqs.~\eqref{eq:Ki} and \eqref{eq:Ii} using \textsc{sagemath} \cite{sagemath}. Additionally, for all $M_x^{(i)}\in  \mathcal{J}_i$, we compute their hyperplane representation to facilitate a direct calculation of $Q^{(i)}$. The explicit results are provided in Appendix \ref{app:Qparam}.

\section{Examples\label{sec:example}}

In this section, we apply our toolbox of spin-adapted GPCs to two few-electron systems: the Lithium and Beryllium atoms. This analysis not only demonstrates the utility of our framework but also reveals that the quasipinning previously observed in these systems \cite{SAKL18} is largely a consequence of spin symmetries.

Before proceeding with the quasipinning analysis, we first outline the methodology used to quantify quasipinning.

\subsection{Quantifying quasipinning and the concept of truncation}\label{sec:QPdistances}

To quantify quasipinning by a spin-adapted GPC $D \geq 0$ one may consider its residual value $D(\bl^{(l)})$, and analogously $D(\bl)$ for a spin-independent GPC. However, this measure has two fundamental shortcomings. First, any GPC $D \geq 0$ can be rescaled according to
\begin{equation}
D(\bl) \geq 0  \quad \Leftrightarrow \quad \widetilde{D}(\bl) \equiv q \,D(\bl) \geq 0\,,
\end{equation}
for any positive scaling factor $q$. Consequently, the residual value $D(\bl^{(l)})$ can be arbitrarily modified without affecting the underlying physics. This ambiguity can be resolved by renormalizing the GPC, for instance, by ensuring that the minimal gap between consecutive eigenvalues of $\hat{D}^{(\Psi)}$ in \eqref{eq:Dhatl} is set to one. Alternatively, one can use the distance of $\bl^{(l)}$ to the respective facet $\mathcal{F}$ of the spectral polytope $\Sigma_{S,M}^{(p)}$, defined by $D \equiv 0$, or to the entire hyperplane extending $\mathcal{F}$ beyond $\Sigma_{S,M}^{(p)\downarrow}$.

Second, since spin-adapted and spin-independent GPCs are only known for relatively small numbers of spatial orbitals, we employ a truncation scheme similar to that in Refs.~\cite{TEVS16, SAKL18}. In this approach, natural (orbital) occupation numbers that are close to zero are neglected, effectively reducing the one-particle Hilbert space $\mathcal{H}_1^{(l)}$ to a subspace of dimension $d'$, for which the GPCs are already known. This introduces a small truncation error \cite{TEVS16, SAKL18}, quantified as the sum of the neglected natural (orbital) occupation numbers:
\begin{equation}
	\varepsilon' := \sum_{j=d'+1}^{d} \lambda_j^{(l)} = N -\sum_{j=1}^{d'} \lambda_j^{(l)} \,.
\end{equation}

To obtain a reliable estimate of quasipinning, we first compute the distance of the truncated vector $\bl^{(l)\prime}$ to any GPC-facet $\mathcal{F}_i^{(p)\prime}$ in the truncated setting $(N,d',S,M)$ and then determine its minimum over all GPCs,
\begin{equation}
F'_{\mathrm{min}} := \min_{i} \mbox{dist}_1(\bl^{(l)\prime},\mathcal{F}_i^{(p)\prime}) \,.
\end{equation}
While $F'_{\mathrm{min}}$ could be interpreted as the minimal distance of $\bl^{(l)\prime}$ to the boundary of $\Sigma_{S,M}^{(p)\prime}$, it is crucial to note that $\bl^{(l)\prime}$ does not, in general, lie within the polytope $\Sigma_{S,M}^{(p)\prime}$ due to its small deviation $\varepsilon'$ from exact normalization.

As shown in Appendix \ref{app:truncation}, this framework allows us to rigorously establish lower and upper bounds on the actual distance of $\bl^{(l)}$ to the boundary of the unknown spectral polytope $\Sigma_{S,M}^{(p)}$ according to
\begin{equation}\label{eq:boundsFmin}
F_{-} \leq \mbox{dist}_1(\bl^{(l)},\partial \Sigma_{S,M}^{(p)}) \leq F_{+}\,,
\end{equation}
where the upper bound is given by $F_{+} \equiv F'_{\mathrm{min}} + \varepsilon'$, while the procedure for computing the lower bound $F_{-} \equiv F_{-}(\bl^{(l)})$ is detailed in the appendix.

\subsection{Lithium and Beryllium atoms}\label{sec:LiandBe}

To obtain accurate approximations of the quantum states for our atomic systems, we performed highly accurate Full Configuration Interaction (FCI) computations of the lowest-energy \textsuperscript{2}S Li and \textsuperscript{3}P Be states using the \textsc{lucita} module \cite{OJS1990,Knecht2008} of the \textsc{dalton2020.1} program package \cite{daltonpaper,dalton2020.1,Note2}.
\footnotetext[2]{To facilitate FCI calculations with large orbital spaces, the constant \texttt{MXTSOB} in the source file \texttt{lucita/mxpdim.inc} must be manually increased before compilation.}
A tight convergence threshold of $10^{-7}$ was applied to the iterative Davidson procedure. Facet distances were computed using the \textsc{clarabel} solver \cite{Clarabel_2024} via the \textsc{cvxpy} package \cite{cvxpy}.

\begin{figure}[tb]
	\includegraphics[width=\linewidth]{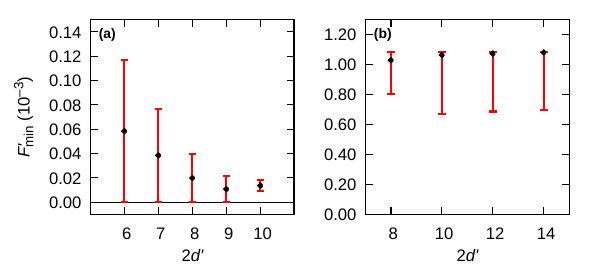}
	\caption{Comparison of the pinning analysis for the Beryllium triplet ground state using
        (a) spin-independent GPCs and (b) spin-dependent GPCs. The point corresponds to the facet
        distance $F'_{\mathrm{min}}$ in the truncated setting, whereas the error bars indicate upper and lower bounds
        defined in Eq.~\eqref{eq:boundsFmin}.\label{fig:Be}}
\end{figure}

For Beryllium, to relate our results to those in \cite{K09} and \cite{SAKL18}, we computed the lowest-energy triplet state with $S=M=1$ for $(N, d) = (4, 25)$ using the ``Be triplet 5'' basis set from Ref.~\cite{SAKL18}, consisting of 25 s-type functions. To compare our spin-adapted GPC analysis with the spin-independent case from Ref.~\cite{SAKL18}, we considered truncated one-particle Hilbert spaces with $2d' \in \{8, 10, 12, 14\}$ for spin-adapted GPCs and $2d' \in \{6, \dots, 10\}$ for spin-independent GPCs. In both cases, we determined the natural occupation number vector $\bd\lambda$ and the natural orbital occupation number vector $\bd\lambda^{(l)}$ of the variational ground state $\ket{\Psi}$. The two panels of Fig.~\ref{fig:Be} show the minimal distance $F_{\mathrm{min}}^\prime$ to the facets of the spin-adapted (right) and spin-independent (left) GPC polytope for different truncated settings $(N, 2d')$. The red error bars indicate the lower and upper bounds given in Eq.~\eqref{eq:boundsFmin}.

\begin{figure}[tb]
	\includegraphics[width=\linewidth]{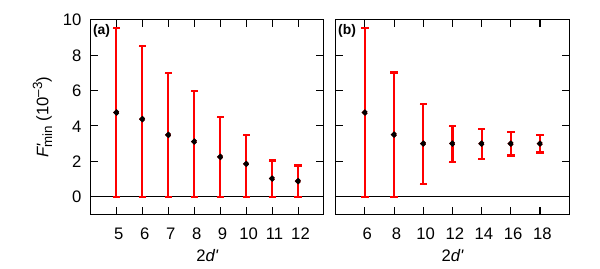}
	\caption{Comparison of the pinning analysis for the Lithium doublet ground state using
        (a) spin-independent GPCs and (b) spin-dependent GPCs. The point corresponds to the facet
        distance $F'_{\mathrm{min}}$ in the truncated setting, whereas the error bars indicate upper and lower bounds
        defined in Eq.~\eqref{eq:boundsFmin}.\label{fig:Li}}
\end{figure}

Compared to the spin-independent case in the left panel, the minimal facet distance increases by an order of magnitude when using spin-adapted GPCs. Notably, exact pinning can already be ruled out for $2d' = 8$ in the spin-adapted case. Furthermore, we assess the non-triviality of quasipinning based on the (spin-adapted) $Q$-parameter introduced in Eq.~\eqref{eq:Q-global} in Sec.~\ref{sec:Q-param}. The spin-adapted $Q$-parameter is computed using the intersections $M_x$ and constants $c_x$ provided in Tables \ref{tab:441}--\ref{tab:471}. In the spin-independent case, we use the $Q$-parameter definitions and coefficients $c$ from Ref.~\cite{TVS17,Note3}.
\footnotetext[3]{The spin-independent $Q$-parameter in \cite{TVS17} was derived by referring to the affine hulls of the facets $\mathcal{F}^{(p)}$ rather than the facets themselves, as done in the present work for determining the spin-adapted $Q$-parameter.}

For spin-independent GPCs with $(N,2d')=(4,10)$, we find $Q=0.864$, whereas for spin-adapted GPCs with $(N,2d')=(4,14)$, we obtain a lower value of $Q=0.369$. This suggests that in the spin-independent case, some portion of the quasipinning is not implied by the Pauli constraints, whereas in the spin-adapted case, nearly all observed quasipinning is trivial, meaning it originates from spin-adapted Pauli constraints. The reduction in the $Q$-parameter further indicates that the non-trivial part of the quasipinning observed in the spin-independent analysis is largely a consequence of spin symmetries.

A similar (quasi)pinning analysis was performed for the Lithium atom in the $S=1/2$ sector, using the ``Li doublet 3'' basis set from Ref.~\cite{SAKL18}. This corresponds to a decontracted aug-cc-pCVQZ basis set \cite{aug-cc-pcvqz-LiBe} with additional tight functions up to $\ell = 4$ and further augmented by h- and i-type functions. In its Cartesian representation and after eliminating linear dependencies with an overlap eigenvalue threshold of $10^{-7}$, this results in the setting $(N, d) = (3, 495)$. We analyzed various truncated settings for which spin-adapted and spin-independent GPCs are available. The minimal facet distances are shown in Fig.~\ref{fig:Li}. As in Ref.~\cite{SAKL18}, we cannot rule out exact pinning even not for the largest truncated setting of $2d'=12$ in the spin-independent case. Moreover, the global $Q$-parameter for the spin-independent analysis (left panel) with $(N,2d')=(3,10)$ is $Q=0.930$, whereas for spin-adapted GPCs with $(N,2d')=(3,18)$, it is significantly lower at $Q=0.024$. This suggests that the previously observed quasipinning was a direct consequence of spin symmetries.

\section{Conclusions and outlook\label{sec:concl}}

In this work, we established a comprehensive toolbox for studying and applying spin-adapted generalized Pauli constraints (GPCs) in few-electron quantum systems. By leveraging the spin symmetry of realistic $N$-electron wave functions, the underlying one-body pure $N$-representability problem simplifies. This, combined with computational resources, enabled us to calculate GPCs for larger active spaces than previously feasible. Notably, for $N=4$ electrons, we successfully determined the spin-adapted GPCs for systems with $d=7,8$ spatial orbitals, a significant improvement over the spin-independent case, where GPCs have so far been computed only for up to ten spin-orbitals ($d=5$). Furthermore, incorporating spin symmetry drastically reduces the number of GPCs, making their application more feasible for realistic system sizes. For example, in the spin-independent setting, the number of GPCs for $N=3,4,5$ electrons and $d=5$ orbitals reaches several hundred, while for $d=7,8$ in the spin-adapted setting, there are still around ten constraints.

To facilitate further applications in physics and chemistry, we provide a complete list of spin-adapted GPCs as supplemental material, including both newly computed cases $(N,d,S)$ and recalculated results for previously studied settings. In that sense, our work represents a new state-of-the-art reference for spin-adapted GPCs and maximizes the availability of constraints across different system sizes.

Beyond these computational advancements, we rigorously proved a fundamental mathematical result that generalizes a similar finding \cite{CBV17} to spinful systems: whenever a spin-adapted GPC is saturated (pinning), the corresponding $N$-electron wave function assumes a constrained structure. More specifically, in a configuration interaction expansion based on natural orbitals, only specific symmetry-adapted configuration state functions may contribute. Furthermore, we established the stability of this selection rule by demonstrating quantitatively that approximate saturation of a GPC (quasipinning) implies an approximate simplification of the wave function.

We also examined the practical significance of (quasi)pinning and showed that it can often be attributed to weak electron correlation or, more generally, to (quasi)pinning by the simpler (spin-adapted) Pauli constraints. To distinguish nontrivial (quasi)pinning from such trivial cases, we introduced a geometric measure, the symmetry-adapted $Q$-parameter, which precisely quantifies the contrast between GPC-induced (quasi)pinning and that arising from Pauli constraints. Applying our toolbox to two atomic systems --- Lithium in its ground state (doublet) and Beryllium in the lowest energy eigenstate of the triplet sector --- our results confirmed that previously observed quasipinning primarily stems from spin symmetry. This aligns with expectations, as spin symmetry inherently simplifies the configuration interaction expansion and, according to the superselection rules, increases the likelihood of (quasi)pinning.

Our findings suggest several promising directions for future research. In one-particle reduced density matrix functional theory (RDMFT) \cite{GPB05, PG16, Piris14, KSPB16, SKB17, SB18, BM18, Piris19, MPU19, SBM19, Piris21-1, P21, DVR21, LKO22, GBM22, MP22, SNF22, LdCP23, LCS23, SG23, GBM23, GBM24, BWR24, VMSS24, CS24, CG24, YS24}, spin-adapted generalized Pauli constraints (GPCs) are expected to play a central role. In pure-state RDMFT, they define the functional domain, while in the ensemble variant, they still influence the structure of the ensemble functional \cite{Schilling18}. As a direct consequence of our proof of the superselection rule and its stability, we predict that the gradients of the universal interaction functionals in various RDMFT variants will exhibit repulsive divergences on the boundary of the functional domain, extending the concept of diverging exchange forces to spinful systems. Since common RDMFT approximations fail to reproduce such divergences, incorporating spin-adapted GPCs into functional development could significantly enhance accuracy, particularly in strongly correlated systems where conventional approaches often struggle.

Analogous to the spin-independent case \cite{SBLMS20}, solving the spin-adapted one-body $N$-representability problem establishes a complete hierarchy of variational multi-configurational self-consistent field (MCSCF) wave function ans\"atze. Each face of the respective polytope corresponds to a distinct ansatz, forming a systematic hierarchy of increasingly sophisticated wave function approximations. This hierarchy allows for successive improvements in systematically capturing electron correlation to higher and higher degrees, starting from the highest weight, which corresponds to the Hartree--Fock point in the spin-independent setting.

Finally, our work underscores the need for more efficient algorithms for computing GPCs, encouraging the mathematical community to bridge the gap between computational feasibility and the practical demands of physics and quantum chemistry. By providing new theoretical insights and computational tools, we hope this work stimulates further research on the role of generalized Pauli constraints in quantum many-body systems.

\begin{acknowledgments}
	We acknowledge financial support from the German Research Foundation (Grant SCHI 1476/1-1), the
    Munich Center for Quantum Science and Technology (J.L., C.S.) and the International Max Planck
    Research School for Quantum Science and Technology (IMPRS-QST) (J.L.). The project/research is
    also part of the Munich Quantum Valley, which is supported by the Bavarian state government
    with funds from the Hightech Agenda Bayern Plus. C.O.\ acknowledges additional support as
    Max-Planck-Fellow at the Max Planck Institute for Solid State Research (MPI-FKF) Stuttgart.
\end{acknowledgments}

\onecolumngrid

\appendix

\section{Selection rule for configuration states\label{app:selection}}

We prove Eq.~\eqref{eq:Delta-variance}, a crucial step in the derivation of the selection rule for spin-adapted configuration states discussed in Sec.~\ref{sec:selection}. Here, the dot above a symbol (e.g., in $\ket{\dot{\Psi}(t)}$) denotes differentiation with respect to $t$, and the map $\mu(\cdot) := \mbox{Tr}_{\mathcal{H}_1^{(s)}}\left[\mbox{Tr}_{N-1}[\cdot]\right]$ represents a partial trace that first traces out $N-1$ particles and subsequently also the spin degree of freedom of the remaining fermion. By following a similar approach as in Ref.~\cite{CBV17} for the spin-independent case, one obtains
\begin{align}
\frac{\mathrm{d}}{\mathrm{d}t}D(\bd\lambda^{(l)}(t))&= \sum_{k=1}^d\kappa_k \bra{k(t)}\dot{\gamma}_l(t) \ket{k(t)}\nonumber\\
&=\sum_{k=1}^d\kappa_k\Tr_{\mathcal{H}_1^{(l)}}\left[ \ket{k(t)}\!\bra{k(t)} \mu\left(\ket{\Psi(t)}\!\bra{\dot{\Psi}(t)} + \ket{\dot{\Psi}(t)}\!\bra{\Psi(t)} \right)\right]\nonumber\\
&= \sum_{k=1}^d\kappa_k \Tr_N\left[E_{kk}\left(\ket{\Psi(t)}\!\bra{\dot{\Psi}(t)} + \ket{\dot{\Psi}(t)}\!\bra{\Psi(t)} \right)\right]\nonumber\\
&= \Tr_N\left[\hat{D}^{(\Psi(t))}\left(\ket{\Psi(t)}\!\bra{\dot{\Psi}(t)} + \ket{\dot{\Psi}(t)}\!\bra{\Psi(t)} \right)\right]\nonumber\\
&= - 2\left( \bra{\Psi(t)}(\hat{D}^{(\Psi(t))})^2\ket{\Psi(t)} - \bra{\Psi(t)}\hat{D}^{(\Psi(t))}\ket{\Psi(t)}^2\right)= - 2\,\text{Var}_{\Psi(t)}\left(\hat{D}^{(\Psi(t))}\right)\,.
\end{align}
In the first line, we applied first-order perturbation theory to the orbital 1RDM $\gamma_l(t)$ to determine $\lambda_k^{(l)}(t)$ for small times, which assumes that the natural orbital occupation numbers $\lambda_k^{(l)}$ are non-degenerate. The treatment of degenerate cases is mathematically more intricate; therefore, we exclude it here and refer the interested reader to the general framework outlined in Ref.~\cite{MSGLS20}.

\section{Bounds on the truncation error}\label{app:truncation}

In this appendix, we derive the relation \eqref{eq:boundsFmin}, which provides a rigorous justification for the truncation procedure.

Due to the truncation error,
\begin{equation}
	\varepsilon' := \sum_{j=d'+1}^{d} \lambda_j^{(l)} = N - \sum_{j=1}^{d'} \lambda_j^{(l)},
\end{equation}
the truncated occupation number vector $\bd\lambda^{(l)\prime}\equiv (\lambda_j^{(l)})_{j=1}^{d'}$ no longer lies within the hyperplane enforcing proper normalization in the Euclidean space $\RR^{d'}$. This introduces a non-uniqueness issue, since for any GPC $D' \geq 0$ in the truncated setting, one can construct an equivalent constraint
\begin{equation}
	\widetilde{D}'(\bd\lambda^{(l)\prime}; \alpha) := D'(\bd\lambda^{(l)\prime}) + \alpha \left(N - \sum_{j=1}^{d'} \lambda_j^{(l)\prime}\right),
\end{equation}
by incorporating the normalization condition with a tunable parameter $\alpha$. For $\varepsilon' > 0$, the residual $D'(\bd\lambda^{(l)\prime})$ can be arbitrarily adjusted by varying $\alpha$, which geometrically corresponds to a rotation of the $(d'-1)$-dimensional hyperplane $D'(\bd\lambda^{(l)\prime}) = 0$ about its intersection with the normalization plane.

To circumvent this ambiguity, we characterize quasipinning not by the residual but by the distance of $\bd\lambda^{(l)}$ to the facet $\mathcal{F}^{(p)}$. The objective is now to estimate this distance in the full setting $(N,d,S,M)$ using the distance of the truncated vector $\bd\lambda^{(l)\prime}$ to the facets $\mathcal{F}^{(p)\prime}$ in the reduced setting $(N,d',S,M)$. To confirm the validity of this approach, we derive the following bounds. Consider the hyperplane
\begin{equation}
    E_D := \Big\{\bd\lambda^{(l)} \in \RR^{d} \big\vert D(\bd\lambda^{(l)}) = 0 \Bigr\}
\end{equation}
in the full setting, and the corresponding hyperplanes
\begin{align}
	E_D'                     & := \Big\{\bd\lambda^{(l)\prime} \in \RR^{d'} \big\vert D'(\bd\lambda^{(l)\prime}) = 0 \Big\},                     \\
	\widetilde{E}_D'(\alpha) & := \Big\{\bd\lambda^{(l)\prime} \in \RR^{d'} \big\vert \widetilde{D}'(\bd\lambda^{(l)\prime}, \alpha) = 0 \Big\}
\end{align}
in the truncated setting, where
\begin{equation}
    D(\bd\lambda^{(l)}) = D'(\bd\lambda^{(l)\prime}) + \sum_{j=d'+1}^{d} \kappa_j \lambda_j^{(l)}.
\end{equation}
Using results from Ref.~\cite{SAKL18}, we obtain the inequality:
\begin{equation}\label{eq:hyperplane-bounds}
	\dist_1(\bd\lambda^{(l)}, E_D) \geq c \dist_1(\bd\lambda^{(l)\prime}, E_D') - \varepsilon'
\end{equation}
with the constant
\begin{equation}
    c = \frac{\max_{1\leq j\leq d'} |\kappa_j|}{\max_{1\leq j\leq d} |\kappa_j|} \leq 1.
\end{equation}
Formally, knowledge of the GPC in the full setting $(N, d, S, M)$ would be required to determine $c$;
however, as was the case for spin-independent GPCs, an analysis of the known GPCs suggests that
$c = 1$ for sufficiently large $d'$.
Since \eqref{eq:hyperplane-bounds} holds for any rotation of $E_D'$ parameterized by $\alpha$, and given that $\dist_1(\bd\lambda^{(l)}, \mathcal{F}^{(p)}) \geq \dist_1(\bd\lambda^{(l)}, E_D)$, we obtain the lower bound:
\begin{equation}
	\dist_1(\bd\lambda^{(l)}, \mathcal{F}^{(p)}) \geq \max_\alpha \dist_1(\bd\lambda^{(l)\prime}, \widetilde{E}_D'(\alpha)) - \varepsilon' \geq 0.
\end{equation}
For the upper bound, following a similar reasoning of subset relations as in Ref.~\cite{SAKL18}, we arrive at:
\begin{equation}
	\dist_1(\bd\lambda^{(l)}, \mathcal{F}^{(p)}) \leq \dist_1(\bd\lambda^{(l)\prime}, \mathcal{F}^{(p)\prime}) + \varepsilon'.
\end{equation}
Combining these results, we establish the following hierarchy:
\begin{equation}\label{eq:facet-bounds}
	\begin{split}
		0 & \leq \max_\alpha \dist_1(\bd\lambda^{(l)\prime}, \widetilde{E}_D'(\alpha)) - \varepsilon' \\
		  & \leq \dist_1(\bd\lambda^{(l)}, E_D)                                                       \\
		  & \leq \dist_1(\bd\lambda^{(l)}, \mathcal{F}^{(p)})                                         \\
		  & \leq \dist_1(\bd\lambda^{(l)\prime}, \mathcal{F}^{(p)\prime}) + \varepsilon'.
	\end{split}
\end{equation}
This establishes rigorous lower and upper bounds on the true facet distance, in particular \eqref{eq:boundsFmin}, ensuring a controlled truncation scheme in practical applications.

\section{Spin-adapted GPCs\label{app:spin-gpcs}}

In this appendix, we provide a complete list of spin-adapted GPCs for various settings $(N,d,S,M)$ solved so far. The tables below include both the GPCs calculated in Ref.~\cite{AK08, Altunbulak-phd} as well as the new constraints for $N=4, S=1$ and $d=7, 8$ reported in this paper in Table \ref{tab:N4-gpcs}. A few comments are in order about trivial constraints not displayed in the tables in the section. For $S=0$, there are no spin-adapted GPCs beyond the Pauli exclusion principle. Moreover, for $(N,S)=(2, 1/2)$ the only constraints are the Pauli exclusion principle $\lambda_1^{(l)}\leq 2$ and the correct normalization $\sum_{i=1}^d\lambda_i^{(l)}=N$. Therefore, we consider $S\geq 1/2$ and $N\geq 3$ in Tables \ref{tab:N3-gpc}-\ref{tab:N6-S32-gpc}.
Furthermore, the spin-adapted GPCs for $N=3$, $S=1/2$ and $d\geq 3$ in Table \ref{tab:N3-gpc} stabilize according to Theorem 4.1.1 in Ref.~\cite{Altunbulak-phd}. Finally, we recall that the constraints for each setting $(N,d,S,M)$ are independent of $M$, i.e., the following constraints in each table are valid for various values $M\in\{-S,-S+1,\ldots, S-1,S\}$.

\begin{table}[htb]
\begin{tabular}{ |c|c| }
 \hline
\makecell{$d$} & \makecell{GPCs} \\
\hline
$d=2$& $\lambda^{(l)}_1 =2$ \\
&$\lambda^{(l)}_2=1$ \\
\hline
$d\ge3$&$\lambda^{(l)}_1 - \lambda^{(l)}_2\le1 $\\
&$\lambda^{(l)}_2 - \lambda^{(l)}_3\le 1$\\
\hline
\end{tabular}
\caption{Spin-adapted GPCs for $N=3, S=1/2$ and $d=\dim(\mathcal{H}_1^{(l)})$. \label{tab:N3-gpc}}
\end{table}

\begin{table}[htb]
\begin{tabular}{ |c|c| }
 \hline
\makecell{$d$} & \makecell{GPCs} \\
\hline
$d=3$&$\lambda^{(l)}_1 =2$ \\
&$\lambda^{(l)}_2=\lambda^{(l)}_3=1$ \\
\hline
&$\lambda^{(l)}_1 - \lambda^{(l)}_3 \le 1$ \\
$d=4$&$\lambda^{(l)}_2 - \lambda^{(l)}_4 \le 1$ \\
& $2\lambda^{(l)}_1 - \lambda^{(l)}_2 + \lambda^{(l)}_4 \le 3$ \\
&$\lambda^{(l)}_2 + 2\lambda^{(l)}_3 - \lambda^{(l)}_4 \le 3$ \\
\hline
&$\lambda^{(l)}_1 - \lambda^{(l)}_3 - \lambda^{(l)}_5 \le 1$\\
\hline
\end{tabular}
\caption{Spin-adapted GPCs for $N=4, S=1$ and $d=\dim(\mathcal{H}_1^{(l)})\in\{3, ..., 8\}$. \label{tab:N4-gpcs}}
\end{table}

\begin{table}[htb]
\begin{tabular}{ |c|c| }
 \hline
\makecell{$d$} & \makecell{GPCs} \\
\hline
$d=3$&$\lambda^{(l)}_1 = \lambda^{(l)}_2=2$ \\
&$\lambda^{(l)}_3=1$ \\
\hline
&$\lambda^{(l)}_1 \le 2$\\
$d=4$&$\lambda^{(l)}_2 - \lambda^{(l)}_3 \le 1$\\
&$\lambda^{(l)}_3 - \lambda^{(l)}_4 \le 1$\\
\hline
&$\lambda^{(l)}_1 \le 2 $\\
&$\lambda^{(l)}_2 - \lambda^{(l)}_3 - \lambda^{(l)}_5 \le 1$\\
$d=5$&$\lambda^{(l)}_3 - \lambda^{(l)}_4 - \lambda^{(l)}_5 \le 1$\\
&$\lambda^{(l)}_1 + \lambda^{(l)}_3 - \lambda^{(l)}_4 \le 3$\\
&$\lambda^{(l)}_1 + \lambda^{(l)}_2 - \lambda^{(l)}_3 \le 3$\\
\hline
&$\lambda^{(l)}_1 \le 2 $\\
&$\lambda^{(l)}_2 - \lambda^{(l)}_3 - \lambda^{(l)}_5 - \lambda^{(l)}_6 \le 1$\\
$d=6$&$\lambda^{(l)}_3 - \lambda^{(l)}_4 - \lambda^{(l)}_5 - \lambda^{(l)}_6 \le 1$\\
&$\lambda^{(l)}_1 + \lambda^{(l)}_3 - \lambda^{(l)}_4 \le 3$\\
&$\lambda^{(l)}_1 + \lambda^{(l)}_2 - \lambda^{(l)}_3 \le 3$\\
\hline
\end{tabular}
\caption{Spin-adapted GPCs for $N=5, S=1/2$ and $d=\dim(\mathcal{H}_1^{(l)})\in\{3, ..., 6\}$.}
\end{table}

\begin{table}[htb]
\begin{tabular}{ |c|c| }
 \hline
\makecell{$d$} & \makecell{GPCs} \\
\hline
$d=4$& $\lambda^{(l)}_1 = 2$\\
& $\lambda_2^{(l)} = \lambda_3^{(l)} =\lambda_4^{(l)} =1$ \\
\hline
&$\lambda^{(l)}_2-2\lambda^{(l)}_3-\lambda^{(l)}_4 \le-1$\\
&$-\lambda^{(l)}_1+\lambda^{(l)}_3-2\lambda^{(l)}_5\le-1$\\
&$\lambda^{(l)}_3-\lambda^{(l)}_4-\lambda^{(l)}_5\le 0$\\
&$\lambda^{(l)}_2-\lambda^{(l)}_3-\lambda^{(l)}_5\le 0$\\
&$\lambda^{(l)}_1-\lambda^{(l)}_3-\lambda^{(l)}_4\le 0$\\
\hline
\end{tabular}
\caption{Spin-adapted GPCs for $N=5, S=3/2$ and $d=\dim(\mathcal{H}_1^{(l)})\in\{4, 5, 6\}$.}
\end{table}

\begin{table}[htb]
\begin{tabular}{ |c|c| }
 \hline
\makecell{$d$} & \makecell{GPCs} \\
\hline
$d=4$&$\lambda^{(l)}_1 = \lambda_2^{(l)}=2$\\
& $\lambda_3^{(l)} =\lambda_4^{(l)} =1$ \\
\hline
&$\lambda^{(l)}_1\le2$\\
&$2\lambda^{(l)}_1-\lambda^{(l)}_2+\lambda^{(l)}_4-2\lambda^{(l)}_5\le3$\\
&$2\lambda^{(l)}_1+\lambda^{(l)}_2-2\lambda^{(l)}_3-\lambda^{(l)}_4\le3$\\
$d=5$&$\lambda^{(l)}_1+2\lambda^{(l)}_2-\lambda^{(l)}_3-2\lambda^{(l)}_4\le3$\\
&$2\lambda^{(l)}_2+\lambda^{(l)}_3-2\lambda^{(l)}_4-\lambda^{(l)}_5\le3$\\
&$\lambda^{(l)}_2+2\lambda^{(l)}_3-\lambda^{(l)}_4-2\lambda^{(l)}_5\le3$\\
&$\lambda^{(l)}_1+\lambda^{(l)}_3-\lambda^{(l)}_5\le 2$\\
&$\lambda^{(l)}_1+\lambda^{(l)}_2-\lambda^{(l)}_4\le 2$\\
\hline
&$\lambda^{(l)}_1\le2$\\
&$\lambda^{(l)}_1+\lambda^{(l)}_3-\lambda^{(l)}_5-\lambda^{(l)}_6\le3$\\
&$\lambda^{(l)}_1+\lambda^{(l)}_2-\lambda^{(l)}_4-\lambda^{(l)}_6\le3$\\
&$3\lambda^{(l)}_1+\lambda^{(l)}_3+2\lambda^{(l)}_4-\lambda^{(l)}_5-2\lambda^{(l)}_6\le9$\\
&$3\lambda^{(l)}_1+2\lambda^{(l)}_2-\lambda^{(l)}_3+\lambda^{(l)}_5-2\lambda^{(l)}_6\le9$\\
&$3\lambda^{(l)}_1+2\lambda^{(l)}_3+\lambda^{(l)}_4-2\lambda^{(l)}_5-\lambda^{(l)}_6\le9$\\
$d=6$&$3\lambda^{(l)}_1+\lambda^{(l)}_2+2\lambda^{(l)}_3-\lambda^{(l)}_4-2\lambda^{(l)}_5\le9$\\
&$3\lambda^{(l)}_1+2\lambda^{(l)}_2+\lambda^{(l)}_3-2\lambda^{(l)}_4-\lambda^{(l)}_5\le9$\\
&$2\lambda^{(l)}_1+3\lambda^{(l)}_2-\lambda^{(l)}_4+\lambda^{(l)}_5-2\lambda^{(l)}_6\le9$\\
&$\lambda^{(l)}_1+2\lambda^{(l)}_2+3\lambda^{(l)}_3-\lambda^{(l)}_5-2\lambda^{(l)}_6\le9$\\
&$\lambda^{(l)}_1+3\lambda^{(l)}_2+2\lambda^{(l)}_3-\lambda^{(l)}_4-2\lambda^{(l)}_6\le9$\\
&$2\lambda^{(l)}_1+\lambda^{(l)}_2+3\lambda^{(l)}_3-2\lambda^{(l)}_5-\lambda^{(l)}_6\le9$\\
&$2\lambda^{(l)}_1+3\lambda^{(l)}_2+\lambda^{(l)}_3-2\lambda^{(l)}_4-\lambda^{(l)}_6\le9$\\
\hline
\end{tabular}
\caption{Spin-adapted GPCs for $N=6, S=1$ and $d=\dim(\mathcal{H}_1^{(l)})\in\{4, 5, 6\}$.}
\end{table}

\begin{table}[htb]
\begin{tabular}{ |c|c| }
 \hline
\makecell{$d$} & \makecell{GPCs} \\
\hline
$d=5$& $\lambda^{(l)}_1 = \lambda_2^{(l)}=2$\\
& $\lambda_3^{(l)} =\lambda_4^{(l)}=\lambda_5^{(l)} =1$ \\
\hline
&$-2\lambda^{(l)}_1-\lambda^{(l)}_2+\lambda^{(l)}_4-3\lambda^{(l)}_6\le-4$\\
&$\lambda^{(l)}_1-\lambda^{(l)}_4-\lambda^{(l)}_5\le0$\\
&$\lambda^{(l)}_2-\lambda^{(l)}_4-\lambda^{(l)}_6\le0$\\
&$\lambda^{(l)}_3-\lambda^{(l)}_5-\lambda^{(l)}_6\le0$\\
&$-\lambda^{(l)}_1+\lambda^{(l)}_3+\lambda^{(l)}_4-2\lambda^{(l)}_6\le0$\\
&$\lambda^{(l)}_1+\lambda^{(l)}_2-\lambda^{(l)}_3-\lambda^{(l)}_4\le1$\\
&$\lambda^{(l)}_3+\lambda^{(l)}_4-\lambda^{(l)}_5-\lambda^{(l)}_6\le1$\\
&$\lambda^{(l)}_1+\lambda^{(l)}_2-\lambda^{(l)}_4\le2$\\
&$\lambda^{(l)}_1+\lambda^{(l)}_3-\lambda^{(l)}_5\le2$\\
&$\lambda^{(l)}_2+\lambda^{(l)}_3-\lambda^{(l)}_6\le2$\\
$d=6$&$2\lambda^{(l)}_1-\lambda^{(l)}_3-\lambda^{(l)}_4+\lambda^{(l)}_6\le2$\\
&$\lambda^{(l)}_1+2\lambda^{(l)}_2-\lambda^{(l)}_4+\lambda^{(l)}_6\le4$\\
&$\lambda^{(l)}_2+2\lambda^{(l)}_3+\lambda^{(l)}_4-\lambda^{(l)}_5\le4$\\
&$2\lambda^{(l)}_1+\lambda^{(l)}_3-\lambda^{(l)}_4+\lambda^{(l)}_6\le4$\\
&$\lambda^{(l)}_2+2\lambda^{(l)}_3+\lambda^{(l)}_5-\lambda^{(l)}_6\le4$\\
&$\lambda^{(l)}_1+2\lambda^{(l)}_2-\lambda^{(l)}_3+\lambda^{(l)}_5\le4$\\
&$\lambda^{(l)}_1+\lambda^{(l)}_3+2\lambda^{(l)}_4-\lambda^{(l)}_5\le4$\\
&$\lambda^{(l)}_2+\lambda^{(l)}_3+2\lambda^{(l)}_4-\lambda^{(l)}_6\le4$\\
&$2\lambda^{(l)}_1+\lambda^{(l)}_2-\lambda^{(l)}_3+\lambda^{(l)}_6\le4$\\
&$3\lambda^{(l)}_1-2\lambda^{(l)}_2-\lambda^{(l)}_3+\lambda^{(l)}_5+2\lambda^{(l)}_6\le4$\\
&$-\lambda^{(l)}_1+\lambda^{(l)}_3+2\lambda^{(l)}_4+3\lambda^{(l)}_5-2\lambda^{(l)}_6\le4$\\
&$3\lambda^{(l)}_1-\lambda^{(l)}_3+\lambda^{(l)}_5+2\lambda^{(l)}_6\le6$\\
\hline
\end{tabular}
\caption{Spin-adapted GPCs for $N=7, S=3/2$ and $d=\dim(\mathcal{H}_1^{(l)})\in\{5, 6\}$. \label{tab:N6-S32-gpc}}
\end{table}

\clearpage

\section{Constants and intersection for calculation of spin-adapted $Q$-parameter \label{app:Qparam}}

In this section, we provide the hyperplane/vertex representations of the intersections $M_x\in \mathcal{J}_i$ \eqref{eq:Ki} and the corresponding parameters $c_x$ for various spin-adapted GPCs and different $N,d,S$ (see also Refs.~\cite{AK08, Altunbulak-phd} for a partial list of GPCs).

\LTcapwidth=\textwidth

\begin{longtable*}[h!]{ | c || c | c | c | c | c | c | c | c |}
\hline
 GPC & $\kappa_0$ & $\kappa_1$ & $\kappa_2$ & $\kappa_3$ &A) V-repr.~of $M_x$ & A) $c$ & B) V-repr.~of $M_x$ & B) $c$\\
 \hline
 1 & 1& -1& 1&0 & \makecell{$\bd v=(2,1,0)$} &  \makecell{$3/4$}& \makecell{same as in A) } &  \makecell{$3/4$}\\
 \hline
  2 & 1& 0& -1& 1& \makecell{same as for GPC 1\\ due to symmetry} & \makecell{same as for GPC 1\\ due to symmetry} & \makecell{same as in A)} & \makecell{$3/4$}\\
  \hline
\caption{Vertex representations (V-repr.) for the intersections $M_x$ and corresponding constants $c_x^{(i)}$ for $(N, d, S)=(3, 3, 1/2)$ and the two settings A) and B). The respective face of the Pauli polytope is zero-dimensional. The $l^1$ distance is used for the calculation of $c_x^{(i)}$.}
\end{longtable*}

\begin{longtable*}[h!]{ | c || c | c | c | c | c | c | c | c | c |}
\hline
 GPC & $\kappa_0$ & $\kappa_1$ & $\kappa_2$ & $\kappa_3$ &  $\kappa_4$ &A) H-repr.~of $M_x^{(i)}$ & A) $c_x^{(i)}$ & B) H-repr.~of $M_x^{(i)}$ & B) $c_x^{(i)}$ \\
\hline
 1 & 1& -1& 1& 0 &0  & \makecell{$(0, 0, 0, 1) \cdot\bd\lambda^{(l)}  = 0$,\\$(2, 0, 1, 0) \cdot\bd\lambda^{(l)} - 4 = 0$,\\$(1, -1, 0, 0) \cdot\bd\lambda^{(l)} - 1 = 0$,\\$(3, 0, 0, 0) \cdot\bd\lambda^{(l)} - 5 \geq 0$,\\$(-1, 0, 0, 0) \cdot\bd\lambda^{(l)} + 2 \geq 0$} &\makecell{1}& \makecell{same as A)} &\makecell{1} \\
 \hline
  2 & 1& 0& -1& 1& 0 & \makecell{$(0, 0, 0, 1) \cdot\bd\lambda^{(l)}  = 0$,\\$(1, 0, 2, 0) \cdot\bd\lambda^{(l)} - 2 = 0$,\\$(1, 2, 0, 0) \cdot\bd\lambda^{(l)} - 4 = 0$,\\$(3, 0, 0, 0) \cdot\bd\lambda^{(l)} - 4 \geq 0$,\\$(-1, 0, 0, 0)\cdot\bd\lambda^{(l)} + 2 \geq 0$} &\makecell{1} &\makecell{same as A)} & \makecell{1}\\
  \hline
\caption{Setting $(N, d, S)=(3, 4, 1/2) $. The inequalities give the hyperplane representation (H-repr.) of the intersection of the boundary of the respective Pauli polytope for A) (spin-adapted) and B) (not spin-adapted) and the respective GPC facet $\mathcal{F}_i^{(p)}$.}
\end{longtable*}

\begin{longtable*}[h!]{ | c || c | c | c | c | c | c | c | c | c | c |}
\hline
 GPC & $\kappa_0$ & $\kappa_1$ & $\kappa_2$ & $\kappa_3$ &  $\kappa_4$ &$\kappa_5$  &A) H-repr.~of $M_x^{(i)}$ & A) $c_x^{(i)}$ & B) H-repr.~of $M_x^{(i)}$ & B) $c_x^{(i)}$  \\
 \hline
 1 & 1& -1& 1& 0 &0&0 &\makecell{$(0, 0, 0, 0, 1) \cdot\bd\lambda^{(l)} = 0$,\\$(2, 0, 1, 1, 0) \cdot\bd\lambda^{(l)} - 4 = 0$,\\$(1, -1, 0, 0, 0) \cdot\bd\lambda^{(l)} - 1 = 0$,\\$(1, 0, -1, 0, 0) \cdot\bd\lambda^{(l)} - 1 \geq 0$,\\$(-2, 0, -1, 0, 0) \cdot\bd\lambda^{(l)} + 4 \geq 0$,\\$(1, 0, 1, 0, 0) \cdot\bd\lambda^{(l)} - 2 \geq 0$} & \makecell{1} & \makecell{\makecell{same as A)}} &\makecell{1} \\
 \hline
  2 & 1& 0& -1& 1& 0&0  &\makecell{$(0, 0, 0, 0, 1)\cdot\bd\lambda^{(l)} = 0$,\\$(1, 2, 0, 1, 0) \cdot\bd\lambda^{(l)} - 4 = 0$,\\$(0, 1, -1, 0, 0)\cdot\bd\lambda^{(l)} - 1 = 0$,\\$(1, -1, 0, 0, 0) \cdot\bd\lambda^{(l)}  \geq 0$,\\$(-1, -2, 0, 0, 0) \cdot\bd\lambda^{(l)} + 4 \geq 0$,\\$(1, 3, 0, 0, 0)\cdot\bd\lambda^{(l)} - 5 \geq 0$} & \makecell{1} & \makecell{ \makecell{same as A)}} &\makecell{1}\\
  \hline
\caption{Setting $(N, d, S)=(3, 5, 1/2) $. The inequalities give the hyperplane representation of the intersection of the boundary of the respective Pauli polytope for A) (spin-adapted) and B) (not spin-adapted) and the respective GPC facet $\mathcal{F}_i^{(p)}$.}
\end{longtable*}

\begin{longtable*}[h!]{ | c || c | c | c | c | c | c | c | c | c | c | c |}
\hline
 GPC & $\kappa_0$ & $\kappa_1$ & $\kappa_2$ & $\kappa_3$ &  $\kappa_4$ &$\kappa_5$ &$\kappa_6$ &A) H-repr.~of $M_x^{(i)}$ & A) $c_x^{(i)}$  & B) H-repr.~of $M_x^{(i)}$ & B) $c_x^{(i)}$ \\
 \hline
 1 & 1& -1& 1& 0 &0 &0&0 &\makecell{$(0, 0, 0, 0, 0, 1)  \cdot\bd\lambda^{(l)} = 0$,\\$(2, 0, 1, 1, 1, 0)  \cdot\bd\lambda^{(l)} - 4 =0$,\\$(1, -1, 0, 0, 0, 0)  \cdot\bd\lambda^{(l)} - 1 = 0$,\\$(1, 0, -1, 0, 0, 0)  \cdot\bd\lambda^{(l)} - 1 \geq 0$,\\$(-2, 0, -1, -1, 0, 0)  \cdot\bd\lambda^{(l)} + 4 \geq 0$,\\$(0, 0, 1, -1, 0, 0)  \cdot\bd\lambda^{(l)} \geq 0$,\\$(2, 0, 1, 2, 0, 0)  \cdot\bd\lambda^{(l)} - 4 \geq 0$}  & \makecell{1} &\makecell{same as A)} & \makecell{1}\\
 \hline
  2 & 1& 0& -1& 1& 0&0&0 &\makecell{$(0, 0, 0, 0, 0, 1) \cdot\bd\lambda^{(l)} = 0$,\\$(1, 2, 0, 1, 1, 0) \cdot\bd\lambda^{(l)} - 4 = 0$,\\$(0, 1, -1, 0, 0, 0) \cdot\bd\lambda^{(l)} - 1 = 0$,\\$(1, -1, 0, 0, 0, 0) \cdot\bd\lambda^{(l)} \geq 0$,\\$(-1, -2, 0, -1, 0, 0) \cdot\bd\lambda^{(l)} + 4 \geq 0$,\\$(0, 1, 0, -1, 0, 0) \cdot\bd\lambda^{(l)} - 1 \geq 0$,\\$(1, 2, 0, 2, 0, 0) \cdot\bd\lambda^{(l)} - 4 \geq 0$}  & \makecell{1} &\makecell{same as A)} & \makecell{1}\\
  \hline
\caption{Setting $(N, d, S)=(3, 6, 1/2) $. The inequalities give the hyperplane representation of the intersection of the boundary of the respective Pauli polytope for A) (spin-adapted) and B) (not spin-adapted) and the respective GPC facet $\mathcal{F}_i^{(p)}$.}
\end{longtable*}

\begin{longtable*}[h!]{ | c || c | c | c | c | c | c | c | c | c | c | c | c |}
\hline
 GPC & $\kappa_0$ & $\kappa_1$ & $\kappa_2$ & $\kappa_3$ &  $\kappa_4$ &$\kappa_5$ &$\kappa_6$&$\kappa_7$ &A) H-repr.~of $M_x^{(i)}$ & A) $c_x^{(i)}$ & B) H-repr.~of $M_x^{(i)}$ & B) $c_x^{(i)}$ \\
 \hline
 1 & 1& -1& 1& 0 &0 &0&0&0  & \makecell{$(0, 0, 0, 0, 0, 0, 1) \cdot\bd\lambda^{(l)}  = 0$,\\$(2, 0, 1, 1, 1, 1, 0) \cdot\bd\lambda^{(l)} - 4 = 0$,\\$(1, -1, 0, 0, 0, 0, 0) \cdot\bd\lambda^{(l)} - 1 = 0$,\\$(1, 0, -1, 0, 0, 0, 0) \cdot\bd\lambda^{(l)} - 1 \geq 0$,\\$(-2, 0, -1, -1, -1, 0, 0) \cdot\bd\lambda^{(l)} + 4 \geq 0$,\\$(0, 0, 1, -1, 0, 0, 0) \cdot\bd\lambda^{(l)} \geq 0$,\\$(0, 0, 0, 1, -1, 0, 0) \cdot\bd\lambda^{(l)} \geq 0$,\\$(2, 0, 1, 1, 2, 0, 0) \cdot\bd\lambda^{(l)} - 4 \geq 0$} & \makecell{1} & \makecell{same as A)} & \makecell{1}\\
 \hline
  2 & 1& 0& -1& 1& 0&0&0&0 & \makecell{$(0, 0, 0, 0, 0, 0, 1)\cdot\bd\lambda^{(l)}  = 0$,\\$(1, 2, 0, 1, 1, 1, 0) \cdot\bd\lambda^{(l)} - 4 = 0$,\\$(0, 1, -1, 0, 0, 0, 0) \cdot\bd\lambda^{(l)}- 1 = 0$,\\$(1, -1, 0, 0, 0, 0, 0) \cdot\bd\lambda^{(l)} \geq 0$,\\$(-1, -2, 0, -1, -1, 0, 0) \cdot\bd\lambda^{(l)} + 4 \geq 0$,\\$(0, 1, 0, -1, 0, 0, 0) \cdot\bd\lambda^{(l)} - 1 \geq 0$,\\$(0, 0, 0, 1, -1, 0, 0) \cdot\bd\lambda^{(l)} \geq 0$,\\$(1, 2, 0, 1, 2, 0, 0) \cdot\bd\lambda^{(l)} - 4 \geq 0$} & \makecell{1} & \makecell{same as A)} & \makecell{1}\\
  \hline
\caption{Setting $(N, d, S)=(3, 7, 1/2) $. The inequalities give the hyperplane representation of the intersection of the boundary of the respective Pauli polytope for A) (spin-adapted) and B) (not spin-adapted) and the respective GPC facet $\mathcal{F}_i^{(p)}$.}
\end{longtable*}

\begin{longtable*}[h!]{ | c || c | c | c | c | c | c | c | c | c | c | c | c | c |}
\hline
 GPC & $\kappa_0$ & $\kappa_1$ & $\kappa_2$ & $\kappa_3$ &  $\kappa_4$ &$\kappa_5$ &$\kappa_6$&$\kappa_7$&$\kappa_8$ &A) H-repr.~of $M_x^{(i)}$ & A) $c_x^{(i)}$ & B) H-repr.~of $M_x^{(i)}$ & B) $c_x^{(i)}$ \\
 \hline
 1 & 1& -1& 1& 0 &0 &0&0&0&0  & \makecell{$(0, 0, 0, 0, 0, 0, 0, 1) \cdot\bd\lambda^{(l)} = 0$,\\$(2, 0, 1, 1, 1, 1, 1, 0) \cdot\bd\lambda^{(l)} - 4 = 0$,\\$(1, -1, 0, 0, 0, 0, 0, 0) \cdot\bd\lambda^{(l)} - 1 =0$,\\$(1, 0, -1, 0, 0, 0, 0, 0) \cdot\bd\lambda^{(l)} - 1 \geq 0$,\\$ (-2, 0, -1, -1, -1, -1, 0, 0) \cdot\bd\lambda^{(l)} + 4 \geq 0$,\\$(0, 0, 1, -1, 0, 0, 0, 0) \cdot\bd\lambda^{(l)} \geq 0$,\\$(0, 0, 0, 1, -1, 0, 0, 0) \cdot\bd\lambda^{(l)} \geq 0$,\\$(2, 0, 1, 1, 1, 2, 0, 0) \cdot\bd\lambda^{(l)} - 4\geq 0$,\\$(0, 0, 0, 0, 1, -1, 0, 0) \cdot\bd\lambda^{(l)}  \geq 0$} &  \makecell{1} & \makecell{same as A)} & \makecell{1} \\
 \hline
  2 & 1& 0& -1& 1& 0&0&0&0&0  & \makecell{$(0, 0, 0, 0, 0, 0, 0, 1)  \cdot\bd\lambda^{(l)} = 0$,\\$(1, 2, 0, 1, 1, 1, 1, 0)  \cdot\bd\lambda^{(l)} - 4 = 0$,\\$(0, 1, -1, 0, 0, 0, 0, 0)  \cdot\bd\lambda^{(l)} - 1 = 0$,\\$(1, -1, 0, 0, 0, 0, 0, 0)  \cdot\bd\lambda^{(l)} \geq 0$,\\$(-1, -2, 0, -1, -1, -1, 0, 0)  \cdot\bd\lambda^{(l)} + 4 \geq 0$,\\$(0, 1, 0, -1, 0, 0, 0, 0)  \cdot\bd\lambda^{(l)} - 1 \geq 0$,\\$(0, 0, 0, 1, -1, 0, 0, 0)  \cdot\bd\lambda^{(l)}  \geq 0$,\\$(1, 2, 0, 1, 1, 2, 0, 0)  \cdot\bd\lambda^{(l)} - 4 \geq 0$,\\$(0, 0, 0, 0, 1, -1, 0, 0)  \cdot\bd\lambda^{(l)}  \geq 0$} & \makecell{1} & \makecell{same as A)} & \makecell{1}\\
  \hline
\caption{Setting $(N, d, S)=(3, 8, 1/2) $. The inequalities give the hyperplane representation of the intersection of the boundary of the respective Pauli polytope for A) (spin-adapted) and B) (not spin-adapted) and the respective GPC facet $\mathcal{F}_i^{(p)}$.}
\end{longtable*}

\begin{longtable*}[h!]{ | c || c | c | c | c | c | c | c | c | c | c | c | c | c | c |}
\hline
 GPC & $\kappa_0$ & $\kappa_1$ & $\kappa_2$ & $\kappa_3$ &  $\kappa_4$ &$\kappa_5$ &$\kappa_6$&$\kappa_7$&$\kappa_8$ &$\kappa_9$&A) H-repr.~of $M_x^{(i)}$ & A) $c_x^{(i)}$ & B) H-repr.~of $M_x^{(i)}$ & B) $c_x^{(i)}$ \\
 \hline
 1 & 1& -1& 1& 0 &0 &0&0&0&0&0  &\makecell{$(0, 0, 0, 0, 0, 0, 0, 0, 1) \cdot\bd\lambda^{(l)}  = 0$,\\$(2, 0, 1, 1, 1, 1, 1, 1, 0) \cdot\bd\lambda^{(l)} - 4 = 0$,\\$(1, -1, 0, 0, 0, 0, 0, 0, 0) \cdot\bd\lambda^{(l)} - 1 = 0$,\\$(1, 0, -1, 0, 0, 0, 0, 0, 0) \cdot\bd\lambda^{(l)} - 1 \geq 0$,\\$(-2, 0, -1, -1, -1, -1, -1, 0, 0) \cdot\bd\lambda^{(l)} + 4 \geq 0$,\\$(0, 0, 1, -1, 0, 0, 0, 0, 0) \cdot\bd\lambda^{(l)} \geq  0$,\\$(0, 0, 0, 1, -1, 0, 0, 0, 0) \cdot\bd\lambda^{(l)}  \geq 0$,\\$(2, 0, 1, 1, 1, 1, 2, 0, 0) \cdot\bd\lambda^{(l)} - 4 \geq 0$,\\$(0, 0, 0, 0, 0, 1, -1, 0, 0) \cdot\bd\lambda^{(l)} \geq 0$,\\$(0, 0, 0, 0, 1, -1, 0, 0, 0) \bd \lambda^{(l)} >= 0$} &  \makecell{1} & \makecell{same as A)} &  \makecell{1}\\
 \hline
  2 & 1& 0& -1& 1& 0&0&0&0&0&0 &\makecell{$(0, 0, 0, 0, 0, 0, 0, 0, 1) \cdot\bd\lambda^{(l)} = 0$,\\$(1, 2, 0, 1, 1, 1, 1, 1, 0) \cdot\bd\lambda^{(l)} - 4 = 0$,\\$(0, 1, -1, 0, 0, 0, 0, 0, 0) \cdot\bd\lambda^{(l)} - 1 = 0$,\\$(1, -1, 0, 0, 0, 0, 0, 0, 0) \cdot\bd\lambda^{(l)} \geq 0$,\\$(-1, -2, 0, -1, -1, -1, -1, 0, 0) \cdot\bd\lambda^{(l)} + 4 \geq 0$,\\$(0, 1, 0, -1, 0, 0, 0, 0, 0) \cdot\bd\lambda^{(l)} - 1 \geq 0$,\\$(0, 0, 0, 1, -1, 0, 0, 0, 0) \cdot\bd\lambda^{(l)} \geq 0$,\\$(1, 2, 0, 1, 1, 1, 2, 0, 0) \cdot\bd\lambda^{(l)} - 4 \geq 0$,\\$(0, 0, 0, 0, 0, 1, -1, 0, 0) \cdot\bd\lambda^{(l)} \geq 0$,\\$(0, 0, 0, 0, 1, -1, 0, 0, 0) \cdot\bd\lambda^{(l)} \geq 0$} & \makecell{1} & \makecell{same as A)}  &  \makecell{1}\\
\end{longtable*}

\begin{longtable*}[h!]{ | c || c | c | c | c | c | c | c | c | c |}
\hline
 GPC & $\kappa_0$ & $\kappa_1$ & $\kappa_2$ & $\kappa_3$ &  $\kappa_4$&A) V-repr.~of $M_x^{(i)}$ & A) $c_x^{(i)}$ & B) H-repr.~of $M_x^{(i)}$ & B) $c_x^{(i)}$\\
 \hline
 1 & 3& -2& 1& 0 &-1 & \makecell{$\bd v = (2, 1, 1, 0)$} & \makecell{$7/10$}&\makecell{same as A)} & \makecell{$7/10$}  \\
 \hline
 2 & 3& 0& -1& -2& 1 &  \makecell{$\bd v = (2, 1, 1, 0)$} &\makecell{$7/10$} & \makecell{same as A)} & \makecell{$7/10$} \\
  \hline
3 & 1& -1& 0& 1& 0& \makecell{$(1, 0, 0, 1)  \cdot\bd\lambda^{(l)} - 2 =0$,\\$(1, 0, -1, 0)  \cdot\bd\lambda^{(l)} - 1 = 0$,\\$(1, 1, 0, 0)  \cdot\bd\lambda^{(l)} - 3 = 0$,\\$(2, 0, 0, 0)  \cdot\bd\lambda^{(l)} - 3 \geq 0$,\\$(-1, 0, 0, 0)  \cdot\bd\lambda^{(l)} + 2 \geq 0$}  &\makecell{$1$} & \makecell{$\bd v = (2, 1, 1, 0)$} &\makecell{$3/5$}  \\
  \hline
 4 & 1& 0& -1& 0& 1& \makecell{same as GPC 3}  &\makecell{$1$} & \makecell{$\bd v = (2, 1, 1, 0)$} & \makecell{$3/5$} \\
  \hline
\caption{Setting $(N, d, S)=(4, 4, 1) $. The inequalities give the hyperplane representation of the intersection of the boundary of the respective Pauli polytope for A) (spin-adapted) and B) (not spin-adapted) and the respective GPC facet $\mathcal{F}_i^{(p)}$.\label{tab:441}}
\end{longtable*}

\begin{longtable*}[h!]{ | c || c | c | c | c | c | c | c | c | c | c |}
\hline
 GPC & $\kappa_0$ & $\kappa_1$ & $\kappa_2$ & $\kappa_3$ &  $\kappa_4$&  $\kappa_5$&A) H-repr.~of $M_x^{(i)}$ & A) $c_x^{(i)}$ & B) H-repr.~of $M_x^{(i)}$ & B) $c_x^{(i)}$ \\
 \hline
 1 & 3& 0& -1 & -2 & 1 & 2 & \makecell{$ (0, 0, 0, 0, 1) \cdot\bd\lambda^{(l)} = 0$,\\
  $ (2, 1, 0, 3, 0) \cdot\bd\lambda^{(l)} - 5 = 0$,\\
$ (1, 2, 3, 0, 0) \cdot\bd\lambda^{(l)} - 7 = 0$,\\
$(1, -1, 0, 0, 0) \cdot\bd\lambda^{(l)}  \geq 0$,\\
$ (1, 5, 0, 0, 0) \cdot\bd\lambda^{(l)} - 7 \geq  0$,\\ $ (-1, -2, 0, 0, 0) \cdot\bd\lambda^{(l)} + 4 \geq 0$} & $1$ & \makecell{same as A)}  & $1$\\
 \hline
 2 & 3& -1 & -2 & 1 & 2 & 0 & \makecell{$(0, 0, 0, 0, 1)  \cdot\bd\lambda^{(l)} =0$,\\ $(1, 0, 0, 1, 0)  \cdot\bd\lambda^{(l)} - 2 = 0$,\\ $ (1, 0, -1, 0, 0)  \cdot\bd\lambda^{(l)} - 1 = 0$, \\$(1, 1, 0, 0, 0)  \cdot\bd\lambda^{(l)} - 3 = 0$,\\$(2, 0, 0, 0, 0)  \cdot\bd\lambda^{(l)}  - 3 \geq 0$,\\ $ (-1, 0, 0, 0, 0)  \cdot\bd\lambda^{(l)}  + 2\geq 0$} &$1$ & \makecell{same as A)} &$1$\\
 \hline
 3 & 3& 0 & -2 & -1 & 2 &1 & \makecell{$(0, 0, 0, 0, 1) \cdot\bd\lambda^{(l)} = 0$, \\
$(1, 0, 0, 2, 0) \cdot\bd\lambda^{(l)} - 2 = 0$, \\
$(0, 0, 1, 0, 0) \cdot\bd\lambda^{(l)} - 1 = 0$, \\
$ (1, 2, 0, 0, 0) \cdot\bd\lambda^{(l)} - 4 = 0$, \\
$(3, 0, 0, 0, 0) \cdot\bd\lambda^{(l)} - 4 \geq 0$, \\
$(-1, 0, 0, 0, 0) \cdot\bd\lambda^{(l)} + 2 \geq 0$}  &$1$ & \makecell{same as A)} & $1$ \\
\end{longtable*}

\begin{longtable*}[tb]{ | c || c | c | c | c | c | c | c | c | c | c | c |}
\hline
 GPC & $\kappa_0$ & $\kappa_1$ & $\kappa_2$ & $\kappa_3$ &  $\kappa_4$ &$\kappa_5$ &$\kappa_6$ &A) H-repr.~of $M_x^{(i)}$ & A) $c_x^{(i)}$ & B) H-repr.~of $M_x^{(i)}$ & B) $c_x^{(i)}$\\
 \hline
 1 & 3& -2& 1& 0 & 1 &0&-1  & \makecell{$(0, 0, 0, 0, 0, 1) \cdot\bd\lambda^{(l)}  = 0$,\\$(0, 1, 0, 0, 1, 0) \cdot\bd\lambda^{(l)} - 1 = 0$,\\$(0, 1, 0, 1, 0, 0) \cdot\bd\lambda^{(l)} - 1 = 0$,\\$(0, 1, -1, 0, 0, 0) \cdot\bd\lambda^{(l)}  =0$,\\$(1, 0, 0, 0, 0, 0) \cdot\bd\lambda^{(l)} - 2 = 0$,\\$(0, 2, 0, 0, 0, 0) \cdot\bd\lambda^{(l)} - 1 \geq 0$,\\$(0, -1, 0, 0, 0, 0) \cdot\bd\lambda^{(l)} + 1 \geq 0$}& \makecell{$1$}& \makecell{same as A)}& \makecell{$1$}\\
 \hline
  2 & 3& -2& -1& 2& 1&0&1 & \makecell{$(0, 0, 0, 0, 0, 1) \cdot\bd\lambda^{(l)} = 0$,\\$(3, 2, -1, 0, 1, 0) \cdot\bd\lambda^{(l)} - 7 = 0$,\\$(2, 1, -2, -1, 0, 0) \cdot\bd\lambda^{(l)} - 3 = 0$,\\$(-2, -1, 3, 0, 0, 0) \cdot\bd\lambda + 3 \geq 0$,\\$(5, 3, -3, 0, 0, 0) \cdot\bd\lambda^{(l)} - 10 \geq 0$,\\$(-2, -1, 1, 0, 0, 0) \cdot\bd\lambda^{(l)} + 4 \geq 0$,\\$(1, 0, -1, 0, 0, 0) \cdot\bd\lambda^{(l)} - 1 \geq 0$}& \makecell{1}& \makecell{same as A)}& \makecell{1}\\
  \hline
  3 & 3& -2& 1& 0& -1&2&1 & \makecell{$(0, 0, 0, 0, 0, 1) \cdot\bd\lambda^{(l)} = 0$,\\$(1, -2, -1, 0, -3, 0) \cdot\bd\lambda^{(l)} + 1 = 0$,\\$(4, 1, 2, 3, 0, 0) \cdot\bd\lambda^{(l)} - 11 = 0$,\\$(1, -2, -1, 0, 0, 0) \cdot\bd\lambda^{(l)} + 1 \geq 0$,\\$(0, 1, -1, 0, 0, 0) \cdot\bd\lambda^{(l)} \geq 0$,\\$(4, 1, 5, 0, 0, 0) \cdot\bd\lambda^{(l)} - 11 \geq 0$,\\$(-1, -1, 1, 0, 0, 0) \cdot\bd\lambda^{(l)} + 2 \geq 0$}& \makecell{1}& \makecell{same as A)}& \makecell{1}\\
\end{longtable*}

\begin{longtable*}[tb]{ | c || c | c | c | c | c | c | c | c | c | c | c | c |}
\hline
GPC & $\kappa_0$ & $\kappa_1$ & $\kappa_2$ & $\kappa_3$ &  $\kappa_4$ &$\kappa_5$ &$\kappa_6$&$\kappa_7$ &A) H-repr.~of $M_x^{(i)}$ & A) $c_x^{(i)}$ & B) H-repr.~of $M_x^{(i)}$ & B) $c_x^{(i)}$ \\
 \hline
 1 & 3& -2& -1& 2 & 1 & 0& 1 &0 & \makecell{$(0, 1, 0, 1, 0, 0, 1) \cdot\bd\lambda^{(l)} - 1 = 0$,\\$(0, 1, 0, 1, 0, 1, 0) \cdot\bd\lambda^{(l)} - 1 = 0$,\\ $(0, 0, 0, 1, -1, 0, 0) \cdot\bd\lambda^{(l)} = 0$,\\$(0, 1, -1, 0, 0, 0, 0) \cdot\bd\lambda^{(l)}  = 0$,\\$(1, 0, 0, 0, 0, 0, 0) \cdot\bd\lambda^{(l)} - 2 = 0$,\\$(0, 1, 0, -1, 0, 0, 0) \cdot\bd\lambda^{(l)}  \geq 0$,\\$(0, -1, 0, -1, 0, 0, 0) \cdot\bd\lambda^{(l)} + 1 \geq 0$,\\$(0, 1, 0, 2, 0, 0, 0) \cdot\bd\lambda^{(l)} - 1 \geq 0$\\\\ \hline \\ $(0, 0, 0, 0, 0, 0, 1) \cdot\bd\lambda^{(l)}  = 0$,\\$(2, 1, -2, -1, 0, -1, 0) \cdot\bd\lambda^{(l)} - 3 = 0$,\\$(3, 2, -1, 0, 1, 0, 0) \cdot\bd\lambda^{(l)} - 7 = 0$,\\$(2, 1, -2, -1, 0, 0, 0) \cdot\bd\lambda^{(l)} - 3 \geq 0$,\\$(0, 0, 1, -1, 0, 0, 0) \cdot\bd\lambda^{(l)}  \geq 0$,\\$(3, 2, -1, 1, 0, 0, 0) \cdot\bd\lambda^{(l)} - 7 \geq 0$,\\$(-2, -1, 1, 0, 0, 0, 0) \cdot\bd\lambda^{(l)} + 4 \geq 0$,\\$(-1, -1, 1, 1, 0, 0, 0) \cdot\bd\lambda^{(l)} + 2 \geq 0$}  & \makecell{$1$\\\\\\\hline\\\\$1$}  & \makecell{same as A)}  & \makecell{$1$\\\\\\\hline\\\\$1$} \\
\end{longtable*}

\begin{longtable*}[h!]{ | c || c | c | c | c | c | c | c | c | c | c | c |}
\hline
 GPC & $\kappa_0$ & $\kappa_1$ & $\kappa_2$ & $\kappa_3$ &  $\kappa_4$ &$\kappa_5$ &A) H-repr.~of $M_x^{(i)}$ & A) $c_x^{(i)}$ & B) H-repr.~of $M_x^{(i)}$ & B) $c_x^{(i)}$\\
 \hline
 1 & 3& -1& 0& -1 & 1 &0 & \makecell{$(0, 1, 2, 0, 1) \cdot\bd\lambda^{(l)} - 4 = 0$,\\$(0, 0, 1, -1, 0) \cdot\bd\lambda^{(l)} - 1 = 0$,\\$(1, 0, 0, 0, 0) \cdot\bd\lambda^{(l)} - 2 = 0$,\\$(0, 1, 3, 0, 0) \cdot\bd\lambda^{(l)} - 5 \geq 0$,\\$(0, 1, -1, 0, 0) \cdot\bd\lambda^{(l)} \geq 0$,\\$(0, -1, -2, 0, 0) \cdot\bd\lambda^{(l)} + 4 \geq 0$} & \makecell{$1$}& \makecell{same as in A)}& \makecell{$1$}\\
 \hline
2 & 3 & -1& -1& 1& 0&0 & \makecell{$(0, 2, 0, 1, 1) \cdot\bd\lambda^{(l)} - 4 = 0$,\\$(0, 1, -1, 0, 0) \cdot\bd\lambda^{(l)} - 1 = 0$,\\$(1, 0, 0, 0, 0) \cdot\bd\lambda^{(l)} - 2 = 0$,\\$(0, 1, 0, 1, 0) \cdot\bd\lambda^{(l)} - 2 \geq 0$,\\$(0, -2, 0, -1, 0) \cdot\bd\lambda^{(l)} + 4 \geq 0$,\\$(0, 1, 0, -1, 0) \cdot\bd\lambda^{(l)} - 1 \geq 0$} & \makecell{$1$}& \makecell{same as in A)}& \makecell{$1$}\\
  \hline
3 & 2 & -1& 0& 0& 0&0 & \makecell{$(0, 1, 1, 1, 1)  \cdot\bd\lambda^{(l)} - 3 = 0$,\\$(1, 0, 0, 0, 0)  \cdot\bd\lambda^{(l)} - 2 = 0$,\\$(0, 1, 1, 2, 0)  \cdot\bd\lambda^{(l)} - 3 \geq 0$,\\$(0, 1, -1, 0, 0)  \cdot\bd\lambda^{(l)} \geq 0$,\\$(0, 0, 1, -1, 0)  \cdot\bd\lambda^{(l)} \geq 0$,\\$(0, -1, -1, -1, 0)  \cdot\bd\lambda^{(l)} + 3 \geq 0$,\\$(0, -1, 1, 0, 0)  \cdot\bd\lambda^{(l)} + 1 \geq 0$,\\$(0, 0, -1, 1, 0)  \cdot\bd\lambda^{(l)} + 1 \geq 0$} & \makecell{$1$}& \makecell{same as in A)}& \makecell{$1$} \\
\end{longtable*}

\begin{longtable*}[h!]{ | c || c | c | c | c | c | c | c | c | c | c | c |}
\hline
 GPC & $\kappa_0$ & $\kappa_1$ & $\kappa_2$ & $\kappa_3$ &  $\kappa_4$ &$\kappa_5$ &$\kappa_6$ & A) H-repr.~of $M_x^{(i)}$ & A) $c_x^{(i)}$ & B) H-repr.~of $M_x^{(i)}$ & B) $c_x^{(i)}$\\
 \hline
 1 & 3& -1& 0& -1 &1 &0&0 & \makecell{$(0, 0, 0, 0, 0, 1) \cdot\bd\lambda^{(l)} = 0$,\\$(2, 1, 2, 0, 1, 0) \cdot\bd\lambda^{(l)} - 8 = 0$,\\$(1, 0, 1, -1, 0, 0) \cdot\bd\lambda^{(l)} - 3 = 0$,\\$(0, 1, -1, 0, 0, 0) \cdot\bd\lambda^{(l)} \geq 0$,\\$(3, 1, 3, 0, 0, 0) \cdot\bd\lambda^{(l)} - 11 \geq 0$,\\$(-1, 0, 0, 0, 0, 0) \cdot\bd\lambda^{(l)} + 2 \geq 0$,\\$(-1, -1, -2, 0, 0, 0) \cdot\bd\lambda^{(l)} + 6 \geq 0$\\\\ \hline \\$(0, 1, 2, 0, 1, 1) \cdot\bd\lambda^{(l)} - 4 = 0$,\\$(0, 0, 1, -1, 0, 0) \cdot\bd\lambda^{(l)} - 1 = 0$,\\$(1, 0, 0, 0, 0, 0) \cdot\bd\lambda^{(l)} - 2 = 0$,\\$(0, 1, 2, 0, 2, 0) \cdot\bd\lambda^{(l)} - 4 \geq 0$,\\$(0, 1, -1, 0, 0, 0) \cdot\bd\lambda^{(l)} \geq 0$,\\$(0, 0, 1, 0, -1, 0) \cdot\bd\lambda^{(l)} - 1 \geq 0$,\\$(0, -1, -2, 0, -1, 0) \cdot\bd\lambda^{(l)} + 4 \geq 0$} & \makecell{$1$\\\\ \\\\ \hline \\ \\\\ $1$}& \makecell{same as in A)}& \makecell{$1$\\\\ \\\\ \hline \\ \\\\ $1$}\\
\end{longtable*}

\twocolumngrid

\bibliography{Refs}

\end{document}